\documentclass{jfm}
\usepackage{graphicx}
\usepackage{epstopdf, epsfig}
\usepackage{float}
\usepackage{amsmath}
\usepackage{xcolor}
\usepackage{mathtools}
\usepackage{lipsum}
\usepackage{float}
\usepackage{color}
\usepackage{comment}
\usepackage[colorlinks,citecolor = blue, linkcolor=red,hyperindex,CJKbookmarks]{hyperref}
\usepackage{siunitx}

\newcommand{\software}[1]{\textsc{#1}}
\newcommand{\massfrac}{y}
\newcommand{\evap}{\mathrm{Ev}}

\newcommand{\mara}{\mathrm{Ma}}
\newcommand{\apextorim}{\text{A}{\rightarrow}\text{R}}
\newcommand{\rimtoapex}{\text{R}{\rightarrow}\text{A}}
\DeclareMathOperator\artanh{artanh}
\DeclareSIUnit{\wtpercent}{wt\%}
\newcommand{\reviewerA}[1]{\textcolor{black}{#1}}
\newcommand{\reviewerB}[1]{\textcolor{black}{#1}}
\newcommand{\reviewerC}[1]{\textcolor{black}{#1}}
\newcommand{\reviewerD}[1]{\textcolor{black}{#1}}

\shorttitle{Marangoni hysteresis and symmetry breaking in binary droplets}
\shortauthor{D. Rocha, D. Lohse and C. Diddens}

\title{Marangoni flow driven hysteresis and azimuthal symmetry breaking in evaporating binary droplets}

\author{
    Duarte Rocha\aff{1}
    \corresp{\email{d.rocha@utwente.nl}},
    Detlef Lohse\aff{1,2}
    \corresp{\email{d.lohse@utwente.nl}},
    \and Christian Diddens\aff{1}
    \corresp{\email{c.diddens@utwente.nl}}
    }

\affiliation{\aff{1}Physics of Fluids Department, Max-Planck Center Twente for Complex Fluid Dynamics and J. M. Burgers Centre for Fluid Dynamics, University of Twente, P. O. Box 217, 7500AE Enschede, The Netherlands
\aff{2}Max-Planck Institute for Dynamics and Self-Organization, Am Faßberg 17, 37077 Göttingen, Germany
}

\begin{document}

\maketitle

\begin{abstract}
The non-uniform evaporation rate at the liquid-gas interface of binary droplets induces solutal Marangoni flows. 
In glycerol-water mixtures (positive Marangoni number, where the \textit{more} volatile fluid has higher surface tension), these flows stabilise into steady patterns. 
Conversely, in water-ethanol mixtures (negative Marangoni number, where the \textit{less} volatile fluid has higher surface tension), Marangoni instabilities emerge, producing seemingly chaotic flows. 
This behaviour arises from the opposing signs of the Marangoni number.
Perturbations locally reducing surface tension at the interface drive Marangoni flows away from the perturbed region.
\reviewerA{Continuity of the fluid} enforces a return flow, drawing fluid from the bulk towards the interface. 
In mixtures with a negative Marangoni number, preferential evaporation of the lower-surface-tension component leads to a higher concentration of the higher-surface-tension component at the interface as compared to the bulk.
The return flow therefore creates a positive feedback loop, further reducing surface tension in the perturbed region and enhancing the instability.
This study investigates bistable quasi-stationary solutions in evaporating binary droplets with negative Marangoni numbers (e.g.\ water-ethanol) and examines symmetry breaking across a range of Marangoni number and contact angles. 
Bistable \reviewerC{droplets} exhibit hysteresis. 
Remarkably, flat droplets (small contact angles) show instabilities at much lower critical Marangoni numbers than droplets with larger contact angles.
Our numerical simulations reveal that interactions between droplet height profiles and non-uniform evaporation rates trigger azimuthal Marangoni instabilities in flat droplets. 
This geometrically confined instability can even destabilise mixtures with positive Marangoni numbers, particularly for concave liquid-gas interfaces\reviewerC{, as they occur in wells}.    
Finally, through Lyapunov exponent analysis, we confirm the chaotic nature of flows in droplets with a negative Marangoni number.
\end{abstract}


\section{Introduction}\label{sec:introduction}

The flow dynamics within evaporating sessile droplets are critical in controlling deposition patterns, which are highly significant for numerous industrial applications such as biological deposition methods \citep{dugas2005droplet}, spray cooling \citep{kim2007spray} or inkjet printing \citep{lohse2022fundamental}. 
In the simplest case of a single-component droplet evaporating under ambient conditions, the evaporation rate is typically non-uniform along the liquid-gas interface, as long as the contact angle ($\theta$) differs from exactly 90$^\circ$ \citep{deegan1997capillary,deegan2000contact,popov2005evaporative}. 
This position-dependent evaporation rate induces capillary flows. 
The best known of such case is the one with a pinned contact line, often referred to as the ``coffee-stain effect'' \citep{deegan1997capillary,deegan2000contact}.
In that case, the mass loss, in the region of higher evaporation, namely at the rim (for droplets with contact angle lower than $90^\circ$), is compensated by a flow towards the rim, while the shape of the droplet remains a spherical-cap.

In cases where evaporative cooling is significant, surface tension gradients arise due to temperature variations at the liquid-gas interface, leading to thermal Marangoni flows that can even overcome the ``coffee-stain effect'' \citep{hu2006marangoni}. 
These flows are driven by thermal gradients. 
In contrast, concentration differences in evaporating \textit{multicomponent} droplets trigger solutal Marangoni flows, which are usually considerably stronger. 
In fact, \citet{Thomson1855XLIIOC} famously first described this phenomenon in the ``tears of wine'' effect, where the faster evaporation of ethanol at the contact line in a wine cup (essentially water-ethanol) results in concentration gradients, driving the upward movement of fluid along the glass. 
As a result, droplets are formed and grow, ultimately falling due to gravity, and leading to the ``tears of wine'' \citep{loewenthal1931xl, hosoi2001evaporative}.

Water-ethanol mixtures, in particular, display a variety of fascinating phenomena (see review of \citet{lohse2020physicochemical} for a general understanding of physicochemical hydrodynamics of multicomponent systems). 
Among these, there is interfacial instability, which manifests as seemingly chaotic flows with erratic asymmetric convection rolls, driven by either (or both) thermal and solutal gradients. 
Initial observations of interfacial instabilities by \citet{benard1901tourbillons} were interpreted as natural convection, but \citet{pearson1958convection} analytically demonstrated that surface tension gradients could also explain some of those instabilities. 
\citet{sternling1959interfacial} investigated  the nature of the solutal Marangoni effect in detail and demonstrated that the direction of the concentration gradient in the bulk, as well as the viscosity and diffusivity ratios between both phases, are key factors for the dynamics.

\begin{figure}
  \centering
  \includegraphics[width=1\textwidth]{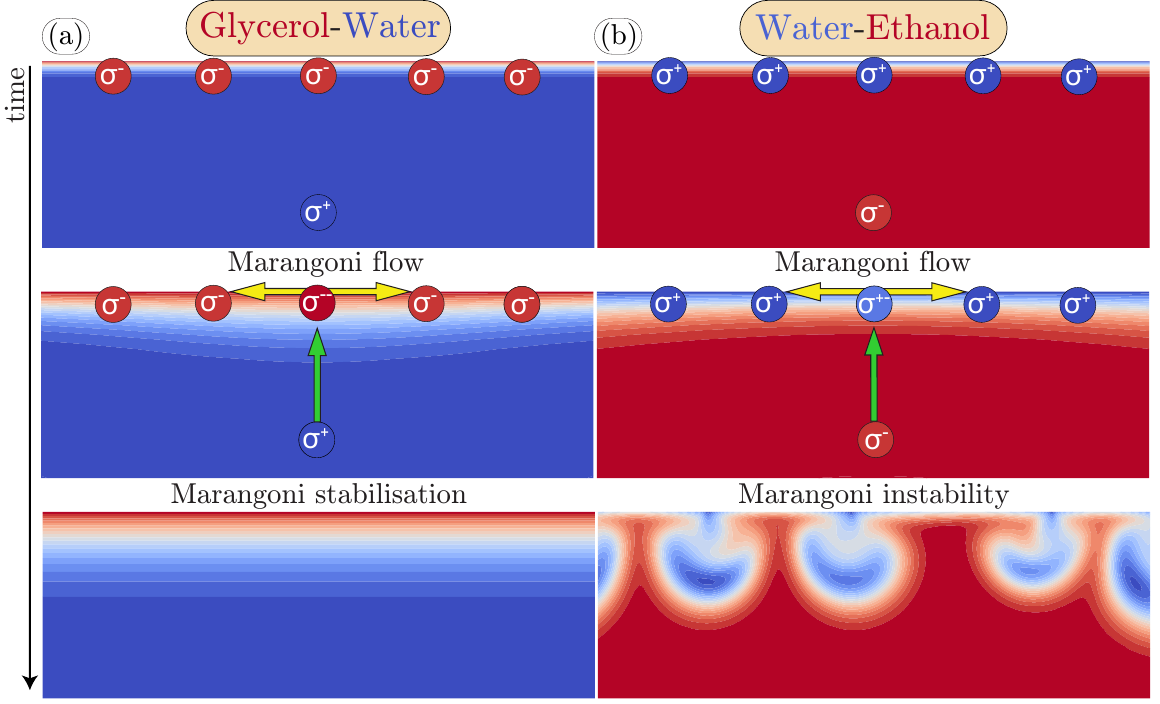}
  \caption{
  Schematic representation of the onset of Marangoni stabilisation (a) and instability (b) in a two-dimensional box with periodic boundaries and with an evaporating top surface, containing a glycerol-water mixture (a) and a water-ethanol mixture (b).
  In (a), due to water being more volatile than glycerol, the surface tension at the interface is reduced as compared to the bulk.
  In (b), vice-versa, due to the higher volatility of ethanol, the surface tension at the interface is enhanced as compared to the bulk.
  A disturbance that locally diminishes surface tension at the interface generates a Marangoni flow directed away from the disturbed area.
  \reviewerA{Continuity of the fluid} generates a return flow from the bulk to the disturbed region.
  In the glycerol-water mixture, this return flow transports liquid with higher surface tension into the disturbed area, thereby mitigating the Marangoni flow.
  Conversely, in the water-ethanol mixture, the return flow carries liquid with lower surface tension into the disturbed area, further reducing the surface tension and amplifying the Marangoni flow.
  Consequently, a positive feedback loop of Marangoni and return flows is established, resulting in the Marangoni instability.
  }
  \label{fig:cartesian_box}
\end{figure}

Interestingly, while these instabilities occur in water-ethanol mixtures, they do not manifest in glycerol-water systems \citep{diddens2021competing}. 
The key difference between these two systems is the sign of the Marangoni number ($\mara$) \citep{gelderblom2022evaporation}.
This difference can be understood by considering two two-dimensional boxes with periodic side boundaries and an evaporating top surface: one containing a glycerol-water mixture (figure \ref{fig:cartesian_box}(a)) and the other containing a water-ethanol mixture (figure \ref{fig:cartesian_box}(b)).
In the glycerol-water system, which has a positive Marangoni number, evaporation of the more volatile component (water) increases the concentration of the fluid with low surface tension (glycerol) at the interface.
Conversely, in the water-ethanol system, which has a negative Marangoni number, evaporation of the more volatile component (ethanol) increases the concentration of the fluid with high surface tension (water) at the interface.
If a perturbation locally reduces surface tension at the interface, it generates a Marangoni flow away from the perturbed region (middle frames in figure \ref{fig:cartesian_box}).
\reviewerA{Continuity of the fluid} enforces a return flow from the bulk to the perturbed region to compensate for the displaced liquid.
In the system with a positive Marangoni number, the return flow brings liquid with higher surface tension into the perturbed region, damping the Marangoni flow.
In contrast, in the system with a negative Marangoni number, the return flow brings liquid with lower surface tension into the perturbed region, further decreasing the surface tension and enhancing the Marangoni flow.
As a result, a positive feedback of Marangoni and return flows is created, leading to the fascinating flow patterns known as Marangoni instability (lower frame in figure \ref{fig:cartesian_box}(b)).

We emphasise that return flow from the bulk towards the interface plays a crucial role in amplifying Marangoni instabilities in this system. 
While the former explanation generally applies to geometries where bulk return flow can develop, in e.g.\ a Hele-Shaw cell \citep{linde1964stromungsuntersuchungen,de2021marangoni}, the influence of alternative geometries on the onset of instabilities remains an open question. 
\reviewerD{Thin-film geometry constraints the development of bulk return flow and Marangoni instabilities arise due to a complex interplay between solutal and thermal Marangoni forces \citep{nazareth2020stability}.}
Droplets have a three-dimensional curved droplet-gas interface, leading to more intricate solutal (and thermal) Marangoni flows compared to simpler geometries, thereby resulting in interesting flows as a product of the complex interaction of its multiple components, see e.g.\ \citet{rowan2000evaporation, sefiane2008wetting,tan2016evaporation,he2016internal, diddens2017evaporating, wang2022wetting,lohse2022fundamental}.
Mixtures with negative Marangoni number exhibit violent flows with larger fluctuations in the early stages, while the concentration of the most volatile component is high, eventually relaxing to a capillary flow when this concentration reduces to a minimal value \citep{christy2011flow, bennacer2014vortices}.
Marangoni instabilities have also been observed in pure ethanol droplets evaporating in a humid environment, where the instabilities were associated with the adsorption of water in the droplet during evaporation \citep{shin2016benard,fukatani2016effect,kita2018quantifying,yang2023evaporation}.
To further complicate the evaporative process, in the later stages of evaporation, the droplet can adopt a pancake-like shape \citep{diddens2017modeling,pahlavan2021evaporation, yang2023evaporation}, where the droplet-gas interface deforms due to Marangoni stresses.
\reviewerD{Previous studies have also documented the occurrence of vigorous convective rolls even in small contact angle water-ethanol droplets \citep[e.g.][]{christy2011flow}, despite the absence of vertical flows.}

In this study, we employ a \reviewerD{\textit{minimal}} model to investigate the flow regimes in quasi-stationary evaporating droplets with negative Marangoni number as a function of the solutal Marangoni number $\mara$ and the contact angle $\theta$, initially under the assumption of axisymmetry. 
We then analyse the azimuthal stability of these solutions for different azimuthal wavenumbers $m$, revealing that flat droplets are more susceptible to instabilities compared to those with larger contact angle $\theta$. 
We utilise a simplified lubrication model to provide scaling arguments, based on the droplet height profile, to justify the emergence of instabilities at low critical $\mara$ number particularly for droplets with low $\theta$. 
Additionally, we show that, counterintuitively, evaporating glycerol-water mixtures can exhibit $\mara$-instabilities when put in a shallow well, illustrating a strong geometric influence on the mechanism of the Marangoni instability.
The influence of droplet deformation on our results is also briefly discussed.
We compute Lyapunov exponents to demonstrate the chaotic nature of these flows using lubrication theory with two lateral dimensions. 
\reviewerD{Throughout this work, we make several simplifications that limit the applicability of the model to real systems.
In particular, our results are only directly applicable to slowly evaporating systems.
Our primary aim is to isolate the fundamental mechanisms driving the onset of Marangoni instabilities and provide a quantitative analysis of the flow stability, which can be refined in future studies to account for more complex and realistic conditions. 
The applicability and limitations of our model are discussed later.
} 

The paper is structured as follows: In \S\ref{sec:physical_model_stokes_droplet}, we introduce the equations used to explore flow regimes in evaporating droplets with negative Marangoni effects, and we introduce the control parameters. 
This is followed by an analysis of quasi-stationary solutions at $\theta$ close to 90$^\circ$ as a function of $\mara$ in \S\ref{sec:instabilities_close_to_90} and the identification of different flow regimes across the full $\mara$-$\theta$ phase space in \S\ref{sec:ma_vs_theta_phase_diagram}. 
The azimuthal stability of the quasi-stationary regimes is studied in \S\ref{sec:ccw_regime} and \S\ref{sec:cw_regime}. 
In \S\ref{sec:lubrication_numerical_method}, we introduce the simplified lubrication model, which is then used to explain the onset of azimuthal instabilities in flat droplets in \S\ref{sec:geometry_azimuthal_instability}. 
In \S\ref{sec:instabilities_glycerol_water}, we propose a well geometry for glycerol-water mixtures exhibiting Marangoni instabilities. 
The influence of droplet deformation is discussed in \S\ref{sec:capillary_effects}.
In \S\ref{sec:chaotic_behaviour}, we compute Lyapunov exponents to illustrate the chaotic behaviour of these flows.
\reviewerD{In \S\ref{sec:limitations}, we discuss the limitations of our model and the implications for real systems.}
The paper ends with the conclusion and outlook (\S\ref{sec:outlook}).

\section{Onset of instability}\label{sec:onset_of_instability}

\subsection{Model and its control parameters}\label{sec:physical_model_stokes_droplet}

Binary droplet evaporation is a complex multi-physics problem involving interfacial mass transfer, which has been successfully analysed numerically using transient \citep{diddens2017evaporating} and quasi-stationary models \citep{diddens2021competing}.  
We study the stability of quasi-stationary flow solutions, focusing on solutal Marangoni instabilities when the most volatile component has lower surface tension. 

\reviewerD{
Typically, rapid evaporation in experiments with water-ethanol droplets induces significant transient and thermal effects.  
Our minimal model isolates solutal Marangoni effects, making it directly applicable only to slowly evaporating systems where thermal effects are negligible.  
Limitations and implications for real systems are discussed in \S\ref{sec:limitations}.}  

When modelling the evolution of each component $\alpha$ in the liquid phase, it is sufficient to track the mass fraction $\massfrac_A$ of the more volatile component $A$, considering that $\massfrac_A + \massfrac_B = 1$. 
Typically, the liquid's velocity $\mathbf{u}$ is much greater than the interface velocity $\mathbf{u}_I$ \citep{diddens2021competing}. 
\reviewerA{For slowly evaporating droplets, we assume only minimal compositional variations in the liquid phase during evaporation, making it reasonable to express $y_A$ as $y_A = y_{A,0} + y$, where $y_{A,0}$ is the spatially averaged composition and $y$ is a small perturbation. 
Naturally, the averaged mass fraction $y_{A,0}$ changes over time; however, under the specified conditions, the process can be considered quasi-stationary at each instant of the drying time, as shown in \S\ref{sec:limitations}.}

The composition-dependent liquid properties can then be approximated using a first-order Taylor expansion around $\massfrac_{A,0}$. 
We neglect the dependence of the liquid's properties on temperature \reviewerD{for simplicity, but we acknowledge these can often be relevant (see \S\ref{sec:limitations}), therefore limiting the model to cases where thermal effects are less pronounced}.
While variations in the liquid density can influence the flow pattern within binary droplets \citep{edwards2018density, li2019gravitational,diddens2021competing}, this effect is neglected in the present study for simplicity, in order to focus instead on pure Marangoni-induced flows. 
The droplet is assumed to perfectly form a spherical-cap shape throughout the calculations in this section, implying that the capillary (and Bond) number(s) are effectively zero.
The contact angle with the substrate is called $\theta$.
All material properties are assumed to be constant regardless of the composition (and temperature), except for the liquid-gas surface tension, given by $\sigma(\massfrac_A) =\sigma(\massfrac_{A,0})+y\partial_{\massfrac_{A}}\sigma$. 
We introduce non-dimensionalised scales (marked with tildes) for space, time, velocity, and pressure:
\begin{alignat*}{4}
  x = V^{1/3}\tilde{x}, &\qquad  t = \dfrac{V^{2/3}}{D_0}\tilde{t}, &\qquad  \mathbf{u} = \dfrac{D_0}{V^{1/3}}\tilde{\mathbf{u}},&\qquad p = \dfrac{\mu_0 D_0}{V^{2/3}}\tilde{p},
\end{alignat*}
\noindent where $V$ is the volume of the droplet and $D_0$ is the diffusion coefficient evaluated at the average composition $\massfrac_{A,0}$.
Similarly, one can obtain the non-dimensionalised gas concentration $\tilde{c}$, whose profile around the droplet determines the evaporation rate of each component, assuming the droplet evaporates under ambient conditions \citep{deegan1997capillary,deegan2000contact,popov2005evaporative}. 
\reviewerC{As demonstrated by \citet{diddens2021competing}, if the droplet is sufficiently well mixed, the influence of local variations in the liquid's composition on the vapour pressure -- and, therefore, on the evaporation rate -- can be neglected.
In other words, effects of Raoult's law are disregarded, and the vapour concentration is assumed to be dependent only on the spatially-averaged composition of the droplet.
This allows us to write the vapour concentration as}
\begin{equation}
\tilde{c} = (c_\alpha - c_\alpha^\infty)/(c^\text{eq}_\alpha - c_\alpha^\infty),
\end{equation}
\noindent where $c_\alpha^\infty$ is the ambient vapour concentration far from the droplet and $c^\text{eq}_\alpha = c_\alpha^\text{pure} \gamma_\alpha x_\alpha$. 
Here, $c_\alpha^\text{pure}$ is the saturation concentration of the pure component $\alpha$ at its average composition, $\gamma_\alpha$ is the activity coefficient, and $x_\alpha$ is the liquid mole fraction.
For simplicity, the saturation concentration $c_\alpha^\text{eq}$ is assumed to be constant.
Following this scaling, $\tilde{c}=1$ at the liquid-gas interface and $\tilde{c}=0$ far from the droplet. 
As originally validated by \citet{deegan2000contact} and \citet{hu2002evaporation} for pure liquids and by \citet{diddens2017detailed} for multicomponent droplets, the solution of the Laplace equation
\begin{equation}
  \nabla^2 \tilde{c} = 0 
  \label{eq:laplace_gas_concentration}
\end{equation}
\noindent in the gas phase gives $\tilde{c}$.
\reviewerC{The solution of equation \eqref{eq:laplace_gas_concentration} depends only on the geometry of the droplet.}
\reviewerA{We disregard Stefan's flow \citep{brutin2018recent} and natural convection in the gas phase based on the results of \citet{diddens2017evaporating}, who have shown that these effects are irrelevant specifically for evaporating water-ethanol droplets at ambient conditions}.
The dimensionless evaporation rate is given by the diffusive mass flux
\begin{equation}
  \tilde{j} = - \tilde{\partial}_n \tilde{c},
  \label{eq:mass_transfer}
\end{equation}
\noindent which induces variations in the liquid composition through the flux boundary condition
\begin{equation}
  -\bnabla \massfrac \bcdot \mathbf{n} = \evap \tilde{j}.
  \label{eq:flux_bc}
\end{equation}
\noindent Here, $\evap$ is the evaporation number, which measures the strength of the concentration gradient created in the liquid due to the preferential evaporation of one of the components, defined as \citep{diddens2021competing}
\begin{equation*}
  \evap = \dfrac{(1 - \massfrac_{A,0})D_A^\mathrm{vap}(c_{A,0}^\mathrm{eq} - c_A^\infty)-\massfrac_{A,0} D_B^\mathrm{vap}(c_{B,0}^\mathrm{eq} - c_B^\infty)}{\rho_0 D_0},
\end{equation*}
In the liquid phase, the conservation of mass and momentum is described by the following Stokes equations, respectively given by
\begin{equation}
  \tilde{\bnabla} \bcdot \tilde{\mathbf{u}} = 0,
  \label{eq:continuity}
\end{equation}
\begin{equation}
  -\tilde{\nabla} \tilde{p} + \tilde{\nabla}^2 \tilde{\mathbf{u}} = 0,
  \label{eq:momentum}
\end{equation}
\noindent where gravity is neglected for simplicity. 
For convenience, we define the modified mass fraction variations $\xi = y / \evap$, which allows us to reduce the number of control parameters of the problem.
When substituting $\massfrac = \evap \xi$ in equation \eqref{eq:flux_bc}, the liquid-gas interface boundary condition becomes
\begin{equation}
  -\bnabla \xi \bcdot \mathbf{n} = \tilde{j}.
  \label{eq:new_flux_bc}
\end{equation}
Since the diffusion coefficient in the liquid phase is typically small, we consider the full convection-diffusion equation for the variations of mass fraction $\xi$
\begin{equation}
  \partial_{\tilde{t}} \xi + \tilde{\mathbf{u}} \bcdot \tilde{\bnabla} \xi = \tilde{\nabla}^2 \xi,
  \label{eq:convection_diffusion}
\end{equation}
Equation \eqref{eq:convection_diffusion} contains the only relevant transient term that can affect the stability of this system, \reviewerB{since the large Schmidt number ($\mathrm{Sc} = \dfrac{\mu_0}{\rho_0 D_0} \sim 10^3$)} indicates that velocity is ``enslaved'' to compositional gradients.
The interface is subject to Marangoni stresses:
\begin{equation}
  \mathbf{n} \bcdot (\tilde{\bnabla} \tilde{\mathbf{u}} + (\tilde{\bnabla}\tilde{\mathbf{u}})^t) \bcdot \mathbf{t} = \mara \tilde{\bnabla}_t \xi \bcdot \mathbf{t},
  \label{eq:marangoni_stress}
\end{equation}
\noindent where $\tilde{\bnabla}_t$ represents the surface gradient operator. 
The Marangoni number $\mara$ is defined as
\begin{equation*}
    \mara = \dfrac{V^{1/3} \partial_{\massfrac_A} \sigma}{D_0 \mu_0} \evap ,
\end{equation*}
\noindent and $\mu_0$ is the dynamic viscosity evaluated at the average composition $\massfrac_{A,0}$.

We consider a spherical-cap for the droplet shape (i.e.\ capillary number $\mathrm{Ca} \rightarrow 0$).
\reviewerD{We restrict our analysis to the case where the droplet is in a quasi-stationary state, such that there is no total mass transfer across the interface.
This scenario can be realised by controlling the ambient humidities of components $A$ and $B$ so that evaporative mass loss of $A$ is compensated by the mass gain of $B$.
In this case, the total mass of the droplet remains constant and thermal gradients are inexistent.
We can therefore enforce $\tilde{\mathbf{u}} \bcdot \mathbf{n} = 0$ at the droplet-gas interface by a normal traction.}
At the substrate, a no-slip boundary condition is imposed, i.e. $\tilde{\mathbf{u}} = 0$, and the Neumann boundary condition $\bnabla \xi \bcdot \mathbf{n} = 0$ is applied to prevent normal mass flux through the substrate.
To eliminate the null space of the pressure field, an average pressure constraint is imposed, i.e., $\int \tilde{p} \, \mathrm{d}V = 0$. 
This is enforced numerically via a Lagrange multiplier. 
Similarly, a zero-average constraint is imposed to remove the constant shift for $\xi$.

\reviewerA{The system of equations is solved using a finite element method implemented in the software package \software{pyoomph}, which is based on \software{oomph-lib} and \software{GiNaC}. 
All domains are discretised using an axisymmetric mesh composed of triangular elements. 
Linear basis functions are employed for $\xi$ and $\tilde{p}$, while quadratic basis functions are used for the $\tilde{\mathbf{u}}$ field. 
The numerical solution process begins with the construction of the residual system of equations, followed by the computation of its Jacobian matrix $\mathbf{J}$ and mass matrix $\mathbf{M}$. 
Quasi-stationary solutions are obtained by providing an initial value for the degrees of freedom and setting the mass matrix contributions to zero. 
Due to hysteresis, the quasi-stationary solutions can vary depending on the initial values assigned to the degrees of freedom. 
After solving for the stationary solution by Newton iteration, the stability of a solution depends on the sign of the real part of the eigenvalues $\lambda$ of the generalised eigenvalue problem $\lambda \mathbf{M} \mathbf{V} + \mathbf{J} \mathbf{V} = 0$, where $\mathbf{V}$ is the eigenvector. 
Stability analysis is conducted using the shift-inverted Arnoldi method to evaluate the eigenvalues. 
If a bifurcation is detected -- where the real part of an eigenvalue crosses zero -- the bifurcation curve is tracked using the method outlined by \citet{diddens2024bifurcation}. 
These bifurcations, classified as either fold or Hopf (in our particular system), lead to distinct types of instabilities. 
Fold bifurcations delimit the range of existence of a particular solution, whereas Hopf bifurcations give rise to oscillatory behaviour, distinguishable by the presence of an imaginary component in the eigenvalue. 
For a more detailed explanation of the numerical methods used, we refer to \citet{diddens2024bifurcation}.
}

\subsection{Instabilities close to $\theta=90^\circ$}\label{sec:instabilities_close_to_90}

In the limit of zero capillary number, there are two control parameters in the problem of this study: $\mara$ and the contact angle $\theta$.
For glycerol-water mixtures, where $\mara$ is positive, the quasi-stationary solutions for the flow and composition fields within the droplet are stable, unique, axisymmetric, and not subject to hysteresis for all $\theta$ \citep{diddens2021competing}.
Exactly at $\theta = 90^\circ$, $\tilde{j}$ is uniform along the liquid-gas interface, resulting in a diffusive profile for $\xi$ and absence of flow within the droplet.
On hydrophilic substrates, i.e. forming contact angles $\theta < 90^\circ$, the larger evaporation rate at the contact line reduces the surface tension locally, inducing a Marangoni flow from the contact line towards the top of the droplet.
On hydrophobic substrates, i.e. forming contact angles $\theta > 90^\circ$, the Marangoni flow is in the opposite direction, since the evaporation rate is larger at the top of the droplet.

In contrast, water-ethanol mixtures, where $\mara$ is negative, exhibit much more complex behaviour.
Strong transient effects and asymmetric, seemingly chaotic flows have been reported both experimentally (\citet{machrafi2010benard,diddens2017evaporating}) and numerically (\citet{diddens2017evaporating}).
Here, we examine the stationary solutions at negative Marangoni numbers $\mara$ with an initially small modulus, before the onset of these instabilities, and gradually decrease $\mara$ to more negative numbers  while monitoring the linear stability of the solutions.

\begin{figure}
  \centering 
  \includegraphics[width=1\textwidth]{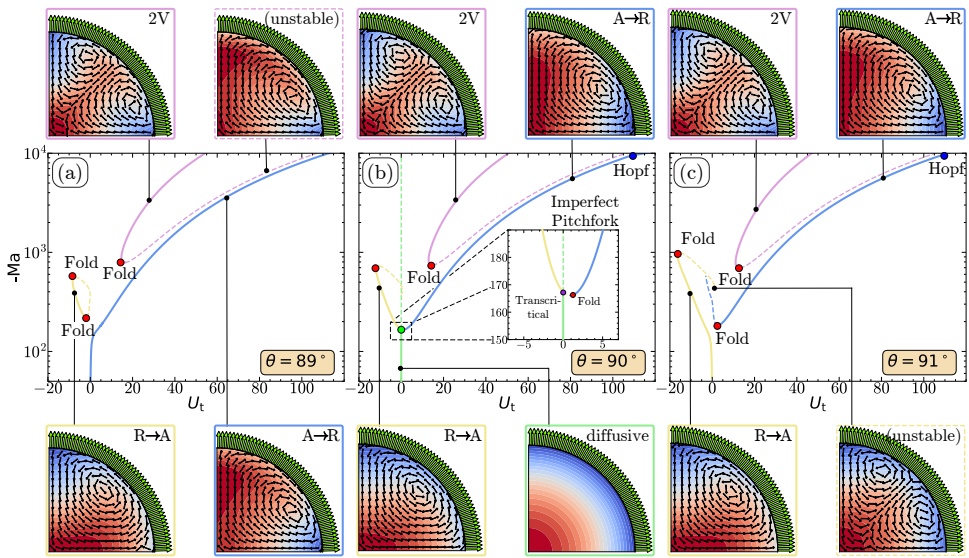}
  \caption{Phase portrait of solution branches as a function of $\mara$, for $\theta = 89^\circ$ (a), $\theta = 90^\circ$ (b), and $\theta = 91^\circ$ (c).
  The $x$-axis represents the average tangential velocity \reviewerA{$U_t$} at the liquid-gas interface, a key measure of the flow field, indicating the flow direction within the droplet when a single vortex is present. 
  The flow field and $\xi$ profile for selected $\apextorim$, $\rimtoapex$, and 2V solutions are also depicted.
  Here, the diverging blue to red colours in the contour plots represent increasing $\xi$ of the fluid with the lowest surface tension, e.g.\ ethanol in water-ethanol mixtures.
  At $\theta = 90^\circ$, the $\xi$ profile is purely diffusive for low $\mara$, becoming unstable at a critical $\mara_\mathrm{cr}$ in an imperfect Pitchfork bifurcation (green).
  The inset in (b) shows the $\theta = 90^\circ$ phase portrait in a range of $\mara$ from 150 to 190, where the imperfect Pitchfork is clearly visible.
  All other solution branches lose their stability in either fold (red) or a Hopf (blue) bifurcations.}
  \label{fig:bif_close_to_90}
\end{figure}

At exactly $\theta = 90^\circ$, we observe a purely diffusive profile for $\xi$, but this solution is only stable at sufficiently low $|\mara|$. 
Here, a non-zero Rayleigh number ($\mathrm{Ra}$), i.e.\ buoyant forces, would initiate flow, but we emphasise again that our analysis here is for $\mathrm{Ra}=0$, i.e.\ negligible density contrasts.
Above the critical $|\mara|$ value, $-\mara>-\mara_\mathrm{cr}\approx 167$, multiple solution branches coexist (see phase portrait in Figure \ref{fig:bif_close_to_90}b). 
Depending on the initial conditions, the flow can develop either a vortex from the apex to the rim ($\apextorim$), a vortex from the rim to the apex ($\rimtoapex$), or two competing vortices (2V). 
\reviewerB{We do not rule out the possibility of additional stable solution branches, such as multiple vortices, but we have not observed them in our axisymmetric numerical simulations.}
At $\mara_\mathrm{cr}$, the flow field's symmetry is broken in an imperfect pitchfork bifurcation (see e.g.\ \citet{golubitsky1979theory,strogatz2018nonlinear}), corresponding to the normal form $\dot{x}=rx+hx^2-x^3$, where $r$ is the control parameter and $h$ the imperfection parameter. 
If $h=0$, we would retrieve a perfect pitchfork bifurcation.
In other words, the diffusive profile becomes unstable through a transcritical bifurcation, where it exchanges stability with the $\rimtoapex$ solution branch, while the $\apextorim$ branch loses stability via a fold bifurcation at a slightly lower $\mara$ (see inset in figure \ref{fig:bif_close_to_90}(b)).
Our numerical tests suggest that the no-slip boundary condition at the substrate and the axisymmetric coordinate system disrupt the physical symmetry that would otherwise be present, leading to the imperfect pitchfork bifurcation. 
At higher $|\mara|$, the $\rimtoapex$, 2V, and $\apextorim$ solution branches each lose stability through either a fold or a Hopf bifurcation.

For slightly lower $\theta$, e.g.\ $\theta = 89^\circ$, the diffusive state no longer exists and the flow exhibits a $\apextorim$ solution for low $|\mara|$.
Despite the induced Marangoni flow pushing the fluid from the top towards the contact line (i.e.\ in the $\apextorim$ direction), a $\rimtoapex$ solution can still be found, which shows strong hysteresis.
The stability of these solutions is limited by fold bifurcations at different $\mara$ (see figure \ref{fig:bif_close_to_90}a).
For slightly larger $\theta$, e.g.\ $\theta = 91^\circ$, the flow exhibits a $\rimtoapex$ solution for low $|\mara|$.
Curiously, this solution branch loses its stability in a fold bifurcation at a lower $|\mara|$ than its $\apextorim$ counterpart for the same set of parameters (see figure \ref{fig:bif_close_to_90}c), which can again be attributed to the three-dimensional geometry and the presence of the no-slip condition.
While Marangoni flow pushes the fluid from the contact line towards the top of the droplet (i.e.\ in the $\rimtoapex$ direction), a $\apextorim$ solution presents, counter-intuitively, for large enough $\mara$, better stability than the $\rimtoapex$ one.

This intriguing introduction to the vast array of bistable solutions within the narrow $\mara$-$\theta$ phase space investigated here sets the stage for the topic to be discussed in the following subsection, which covers the stability of the quasi-stationary solutions on an extensive $\mara$-$\theta$ phase diagram.

\subsection{Ma vs $\theta$ phase diagram}\label{sec:ma_vs_theta_phase_diagram}

\begin{figure}
  \centering
  \includegraphics[width=0.8\textwidth]{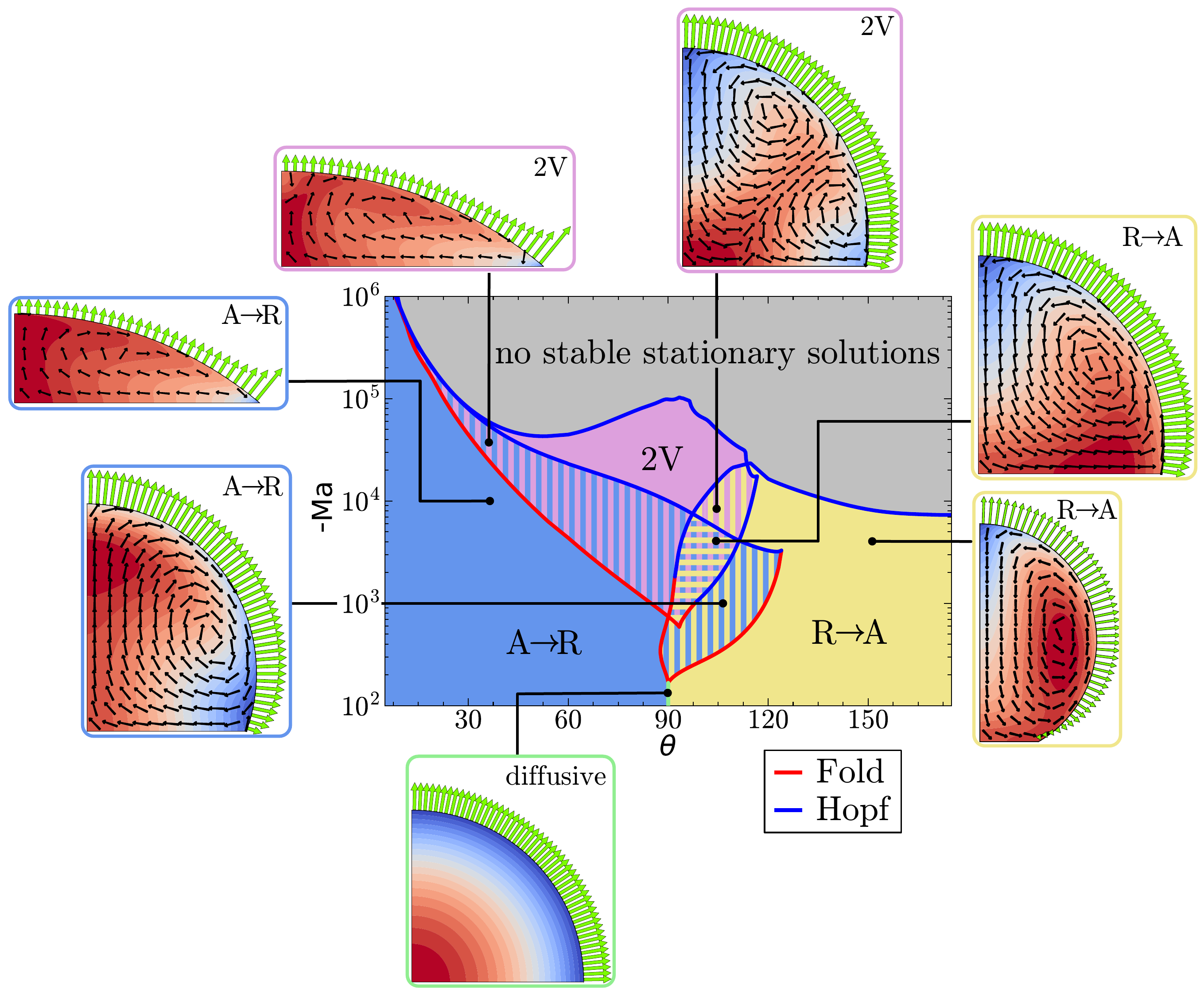}
  \caption{Bifurcation diagram of the quasi-stationary axisymmetric solutions as a function of $\mara$ and $\theta$.
  The diagram is divided into three regimes, where the stable solutions are either a $\apextorim$ solution (blue colour), a $\rimtoapex$ solution (yellow colour), or a 2V solution (plum colour).
  The transition between these regimes is marked by fold (red) or Hopf (blue) bifurcations.
  Above the upper Hopf bifurcation curves, no stable quasi-stationary solutions are found (gray).}
  \label{fig:ma_vs_theta}
\end{figure}

In this section, we utilise the bifurcation tracking tool outlined by \citet{diddens2024bifurcation}. 
Building upon the initial findings from the preceding subsection, we employ arclength continuation on $\theta$ to trace the bifurcation curves along the $\mara$-$\theta$ phase diagram.
Arclength continuation on $\theta$ requires remeshing the domain whenever the grid deformation surpasses a certain threshold.
We used highly refined meshes near the liquid-gas interface, with a total of $\sim$\SI{40000} degrees of freedom, to accurately capture the Marangoni flow. 
The curves in figure \ref{fig:ma_vs_theta} remained unchanged with further grid refinement, confirming mesh independence. 

We find a complex bifurcation diagram where multiple solution branches coexist across a broad spectrum of $\mara$ and $\theta$ values (figure \ref{fig:ma_vs_theta}).
This diagram is segmented into three distinct regimes, each representing stable quasi-stationary axisymmetric $\apextorim$, 2V or $\rimtoapex$ solutions (blue, plum and yellow colours, respectively).
Transitions between these regimes are characterised by fold or Hopf bifurcations, indicating shifts in the stability of solution branches. 
By calculating the oscillation amplitude $A$ of the \textit{rms}-velocity of transient oscillatory solutions near selected Hopf bifurcations, we observe that these ultimately reach a stable limit cycle, $A \sim \sqrt{\mara - \mara_\mathrm{cr}}$. 
This allows us to classify these Hopf bifurcations as supercritical.
Under the assumptions and flow conditions of this subsection, we have not observed any subcritical Hopf bifurcations.

It is important to note the absence of stable quasi-stationary solutions beyond the upper Hopf bifurcation curves. 
In this region, the flow field is dominated by nonlinear effects, which can lead to seemingly chaotic behaviour, as previously observed in both experimental and numerical studies (\citet{machrafi2010benard,diddens2017evaporating}).
\reviewerA{Additionally, the upper Hopf bifurcation curves are not smooth, contrary to the other bifurcation curves. 
This irregularity arises from the coexistence of multiple solution branches that lose stability at different $\mara$ values, resulting in a complex and non-smooth bifurcation structure.}
Interestingly, the bifurcation diagram reveals a network of bistable regions. 
Within these areas, the flow field can adopt solutions from different regimes based on initial conditions, demonstrating significant hysteresis.

The calculated set of solutions provides a comprehensive overview of the stability of the quasi-stationary axisymmetric solutions within binary droplets with negative $\mara$.
However, the azimuthal stability of these solutions has not yet been addressed in this section, which is the focus of the following sections.

\section{Azimuthal stability}\label{sec:azimuthal_stability}

We employ the method detailed in \citet{diddens2024bifurcation} to explore the phenomenon of axisymmetric breaking across the $\rimtoapex$, 2V and $\apextorim$ solution regimes. 
We determine the stability of the quasi-stationary axisymmetric base solutions for specific azimuthal wavenumbers $m$, i.e.\ the stability of perturbations $\propto \exp(im\phi)$, where $\phi$ is the azimuthal coordinate.

\subsection{Regimes with vortices from rim to apex or with two vortices}\label{sec:ccw_regime}

\begin{figure}
  \centering
  \includegraphics[width=1\textwidth]{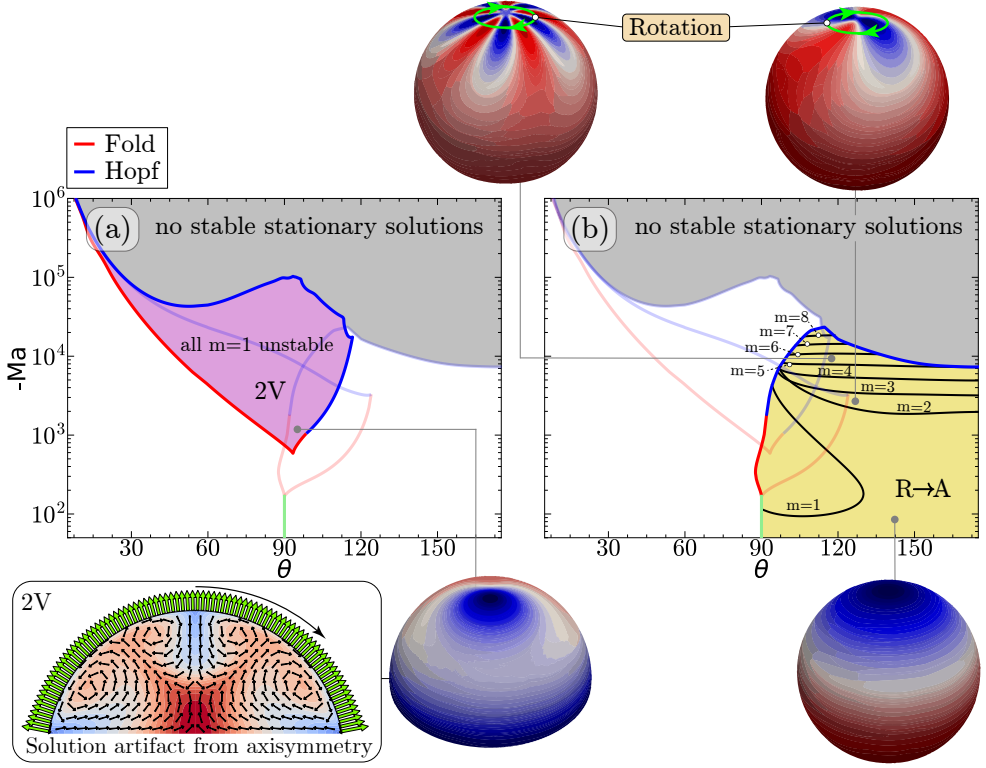}
  \caption{Azimuthal bifurcation diagram for the stability of the quasi-stationary axisymmetric solutions in the 2V (a) and $\rimtoapex$ (b) solutions regimes.
  In (a), the bifurcations that limit the stability of the $\apextorim$ and $\rimtoapex$ solution regimes are shown with lower opacity. 
  Similarly, in (b), the bifurcations for the $\apextorim$ and 2V solutions are depicted with lower opacity (see all curves in figure \ref{fig:ma_vs_theta}).
  In the 2V regime, all base solutions are unstable for $m=1$, leading to a single vortex, as depicted at the bottom right of (a).
  The 2V regime is therefore interpreted as an artifact of the imposed axisymmetry in $\S \ref{sec:ma_vs_theta_phase_diagram}$, where the upper vortex merges with its counterpart in the rim region of the droplet, as depicted at the bottom left of (a).
  In the $\rimtoapex$ regime, multiple bifurcations are observed corresponding to the instability of modes $m=1$ to $m=8$.
  The adjacent plots show an isometric view of an azimuthally stable solution (bottom right), and solutions subject to the linear effects of $m=2$ (top right), and $m=5$ (top left) azimuthally unstable perturbations.
  Exclusively in the $\rimtoapex$ regime, the eigenvalues and eigenfunction had a non-zero imaginary part, indicating rotational motion, as depicted by the green arrows in the two upper plots.
  }
  \label{fig:ccw_azimuthal_stability}
\end{figure}

We first focus on the 2V regime (figure \ref{fig:ccw_azimuthal_stability}a). 
All of these base solutions exhibit an instability for $m = 1$.
\reviewerC{If instabilities are triggered, the flow field will be dominated by a single vortex.
The upper vortex merges with its counterpart in the rim region of the droplet, resulting in a single vortex (see bottom right of figure \ref{fig:ccw_azimuthal_stability}a).
Consequently, the 2V flow field is merely an artifact of the imposed axisymmetry during the computation of the base solutions.}

In contrast, the $\rimtoapex$ vortex regime (see figure \ref{fig:ccw_azimuthal_stability}b) presents a variety of bifurcations, along with a stable axisymmetric regime (e.g.\ at $\theta=140^\circ$ and $-\mara=100$, as shown in the bottom right of figure \ref{fig:ccw_azimuthal_stability}b).
For lower $\theta$ within this regime, a $m=1$ bifurcation at a relatively low $-\mara_\mathrm{cr} \sim 100$ breaks the axisymmetry. 
Interestingly, the axisymmetric stability can be regained by sufficiently increasing $|\mara|$.
However, if $|\mara|$ continues to rise, the base solution becomes unstable for $m=2$, and subsequently for $m=3$ up to $m=8$.
As $|\mara|$ exceeds the critical value for each bifurcation, the axisymmetry can be disrupted in a corresponding mode $m$, as depicted in the adjacent plots of figure \ref{fig:ccw_azimuthal_stability}b.
These plots are derived by expanding the $\xi$ field into a sum of the base solution and a small amplitude perturbation in the direction of the corresponding eigenvector that triggers the instability of mode $m$.
While in the long-term limit the resulting $\xi$ profile will be governed by nonlinear coupling of additional excited modes, we only represent the linear effects of each mode to offer a clear visualisation of the azimuthal instability.
We highlight the presence of a non-zero imaginary part in the eigenvalue and eigenfunction of the velocity field for the $\rimtoapex$ regime, indicating rotational motion in the perturbed flow field, as depicted by the green arrows in the top plots.
This result is consistent with the reports of \citet{babor2023linear} for thermal Marangoni flow in sessile droplets.

\subsection{Regime with vortex from apex to rim}\label{sec:cw_regime}

\begin{figure}
  \centering
  \includegraphics[width=1\textwidth]{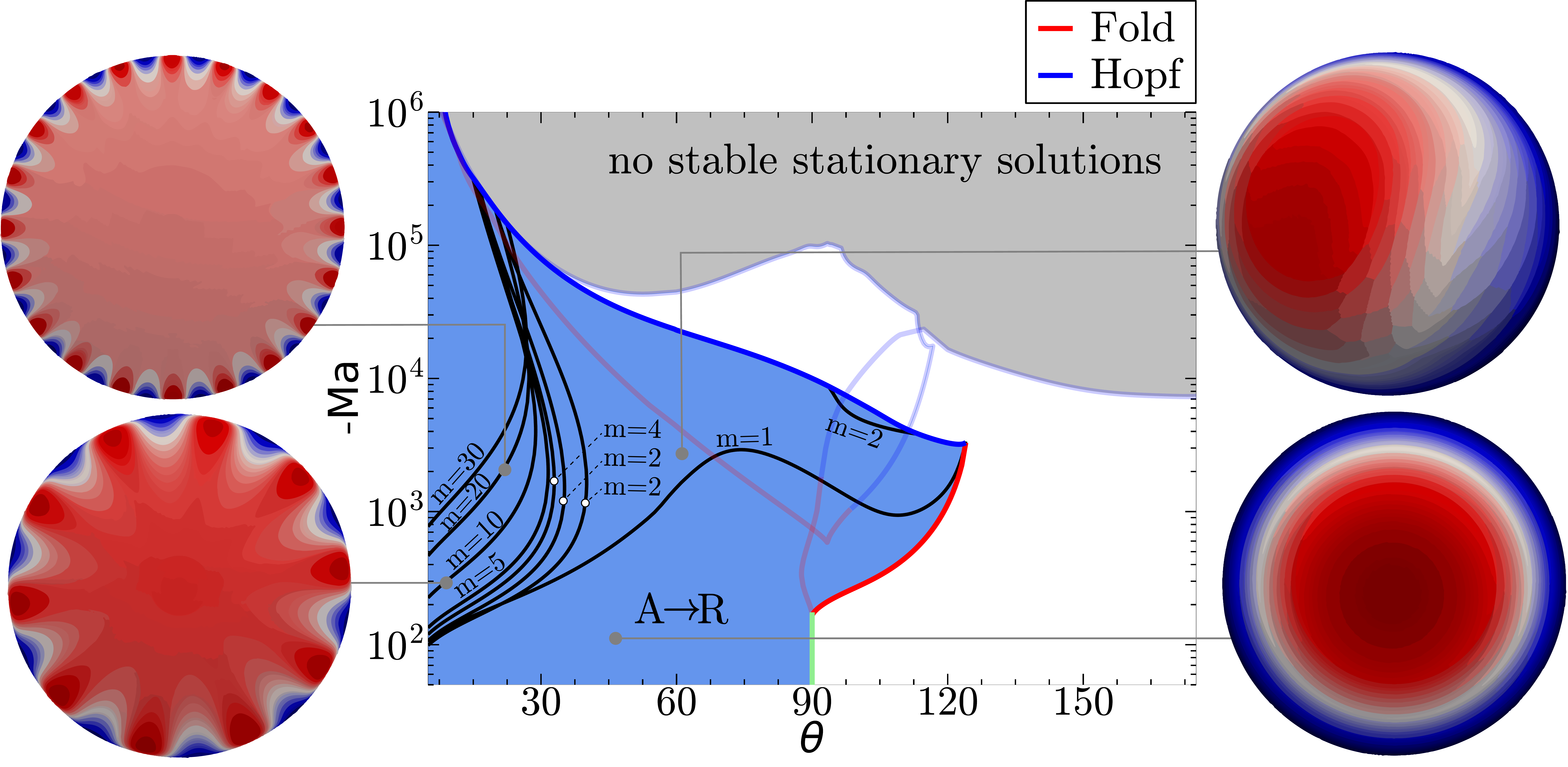}
  \caption{Azimuthal bifurcation diagram assessing the stability of the quasi-stationary axisymmetric solutions in the $\apextorim$ regime.
  The bifurcations that limit the stability of the 2V and $\rimtoapex$ solution regimes are shown with lower opacity (see all curves in figure \ref{fig:ma_vs_theta}). 
  The bifurcation curves depict the range from $m=1$ to $m=30$.
  The adjacent plots show a top view of the droplet in a stable axisymmetric regime (bottom right) or subject to the linear effects of $m=1$ (top right), 10 (bottom left), and 20 (top left) unstable perturbations on the azimuthal instability.}
  \label{fig:cw_azimuthal_stability}
\end{figure}

In the regime of the $\apextorim$ solution, we observe multiple bifurcations for various $m$ values (see figure \ref{fig:cw_azimuthal_stability}). 
For larger $\theta$, greater than $45^\circ$, the axisymmetric base solution remains stable for $|\mara|$ values below a critical $m=1$ bifurcation (see adjacent plot at the bottom right of figure \ref{fig:cw_azimuthal_stability}), while it becomes unstable for $m=1$ above this bifurcation (see adjacent plot at the top right of figure \ref{fig:cw_azimuthal_stability}).
Additionally, there exists a narrow region where certain combinations of $\mara$ and $\theta$ lead to a $m=2$ unstable solution. 
This region is observed for $\theta$ values slightly above $90^\circ$ and $\mara \sim -10^4$.

For lower $\theta$, we observe a multitude of bifurcations, as illustrated in figure \ref{fig:cw_azimuthal_stability}, ranging from $m=1$ to $30$ (instabilities with $m>30$ are not indicated in figure~\ref{fig:cw_azimuthal_stability}). The bottom and top left adjacent plots provide representations for $m=10$ and $20$, respectively. 
Interestingly, the critical Marangoni number $\mara_\mathrm{cr}$ for the onset of all instabilities is particularly low for smaller $\theta$. 
This observation contradicts the explanation given for Marangoni instabilities in a two-dimensional box in section \S\ref{sec:introduction}.
In a two-dimensional box, the instabilities are driven by a positive feedback loop, where a perturbation locally reduces surface tension at the interface, inducing a Marangoni flow. 
This, in turn, generates a return flow in the bulk, drawing more fluid with low surface tension towards the perturbed region.
However, in the case of flat droplets (i.e. small $\theta$), the bulk domain becomes so shallow that vertical diffusion averages out any considerable compositional gradients in height direction, consistent with the assumptions of lubrication theory. 
Thus, the vertical return flow is absent, breaking the positive feedback loop.
Contrary to expectations that Marangoni instabilities would be suppressed for low $\theta$ values, we observe the opposite behaviour.

To deeply understand this intriguing occurrence, we introduce a lubrication model in the following section. 
This model reduces the number of parameters to one, namely just a Marangoni number. 
In the limit of small $\theta$, the contact angle can be absorbed in the nondimensionalisation.

\section{Instabilities for low contact angle $\theta$}\label{sec:lubrication_ca_0}

\subsection{Lubrication model and its control parameter}\label{sec:lubrication_numerical_method}

In the limit of small $\theta$, the vertical gradients are negligible compared to the radial or azimuthal ones.
Rather than solving the full axisymmetric equations, we can then simplify the problem by applying lubrication theory.
This allows us to simply consider a one-dimensional grid using a polar coordinate system.
Alternatively, a two-dimensional circular mesh could be used, taking a Cartesian coordinate system, albeit with an increased computational cost.
We choose the first option in this section.

\begin{figure}
  \centering
  \includegraphics[width=1\textwidth]{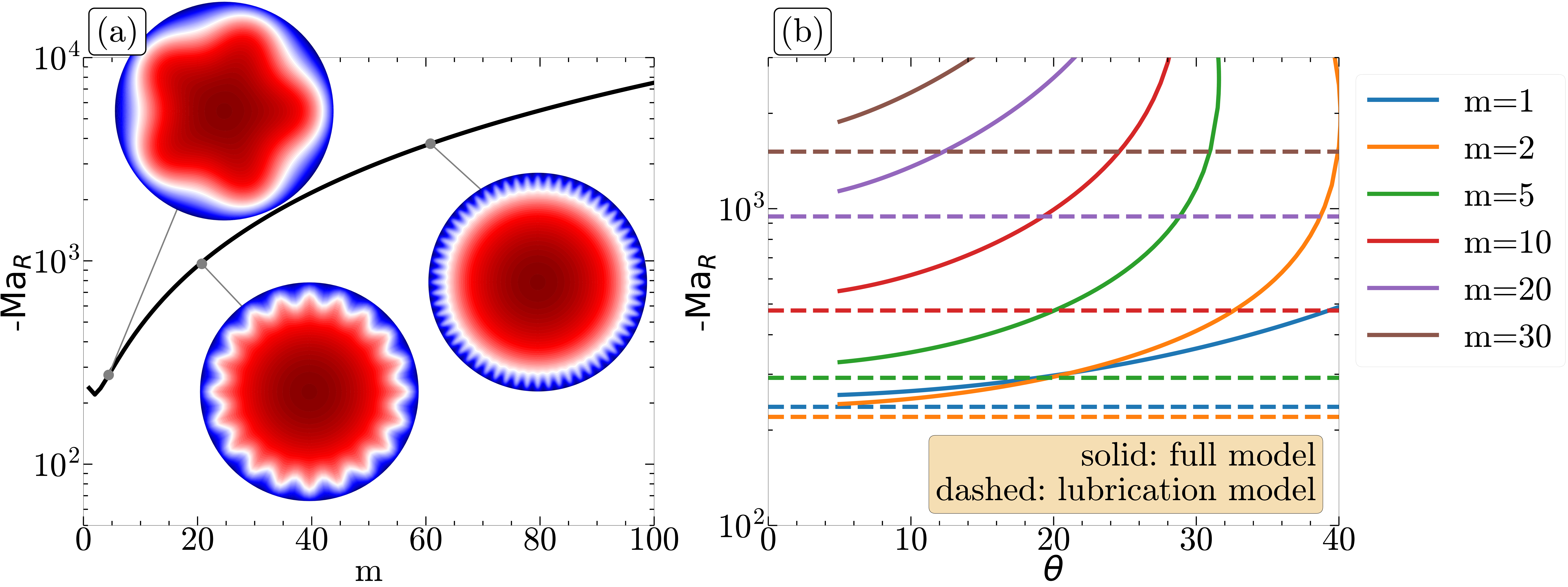}
  \caption{(a) Onset of azimuthal instabilities for different $m$ as a function of $\mara_R$ in the limit of small $\theta$. 
  The adjacent plots show a top view of the droplet subject to the linear effects $m=5$ (top left), $m=20$ (bottom left), and $m=60$ (bottom right) unstable perturbations on the azimuthal instability.
  (b) Comparison of the azimuthal stability of the quasi-stationary axisymmetric solutions between the full model as $\theta \rightarrow 0$ (solid) and the lubrication model (dashed).}
  \label{fig:lubrication_validation}
\end{figure}

In the lubrication approximation of an evaporating droplet \citep{eggers2010nonlocal,diddens2017modeling}, the governing equations for the pressure $p$, height profile $h$ and vertically averaged mass fraction $y_A$ are given by, assuming a lateral flow profile parabolic in height direction:
\
\begin{equation}
  \partial_t h = -\bnabla \bcdot \mathbf{Q} - \dfrac{j}{\rho},
  \label{eq:lubrication_h}
\end{equation}
\begin{equation}
  p + \bnabla \bcdot (\sigma \bnabla h) = 0,
  \label{eq:lubrication_p}
\end{equation}
\begin{equation}
  \partial_t (h y_A) + \nabla \bcdot (\mathbf{Q} y_A) = \bnabla \bcdot (D h \bnabla y_A) - \dfrac{j_A}{\rho},
  \label{eq:lubrication_y}
\end{equation}
\begin{equation}
  \mathbf{Q} = -\dfrac{h^3}{3 \mu} \bnabla p + \dfrac{h^2}{2 \mu} \bnabla \sigma,
  \label{eq:lubrication_Q}
\end{equation}
\
Note that the transient effects are more pronounced for flat droplets as compared to droplets with higher contact angle.
However, we calculate the quasi-stationary solutions of the lubrication model for simplicity.
We assume that the evaporation of one component is balanced by the condensation of the other in the quasi-stationary limit, such that the total volume remains constant.
Thereby, we consider the evaporation rate $j_A$ to be given by the theoretically calculated limit for a flat pure droplet \citep{popov2005evaporative},
\
\begin{equation}
  j_A = \dfrac{2 D_A^\mathrm{vap} (c_A^\mathrm{eq} - c_A^\infty)}{\pi R \sqrt{1 - r^2}} \, .
\end{equation}
\
As in \S\ref{sec:physical_model_stokes_droplet}, we assume that the droplet maintains its initial shape, i.e.\ we consider the limit of zero capillary number $\mathrm{Ca}=0$ and $\partial_t h = 0$. 
In the limit of the lubrication theory, the equilibrium shape corresponds to a parabolic height profile.
We emphasise that these approximations may limit the applicability of the results in real systems, but they capture the essential physics that explain the onset of the Marangoni instability.
By reformulating $\xi = y \theta / \mathrm{Ev}$ and applying the mentioned assumptions, while using the nondimensional scales introduced in \S\ref{sec:physical_model_stokes_droplet} -- but now modifying the length scale to the base radius $R$ instead of $V^{1/3}$ -- we can reduce the system of equations \eqref{eq:lubrication_h}--\eqref{eq:lubrication_Q} to:
\
\begin{equation}
  \tilde{\bnabla} \bcdot \tilde{\mathbf{Q}} = 0,
  \label{eq:lubrication_dimensionless_div_Q}
\end{equation}
\begin{equation}
  \reviewerC{\tilde{h} \partial_{\tilde{t}} \xi + }
  \tilde{\bnabla} \bcdot (\tilde{\mathbf{Q}} \xi) = \tilde{\bnabla} \bcdot (\tilde{h} \tilde{\bnabla} \xi) - \tilde{j},
  \label{eq:lubrication_dimensionless_y}
\end{equation}
\begin{equation}
  \tilde{\mathbf{Q}} = - \dfrac{\tilde{h}^3}{3} \tilde{\bnabla} \tilde{p} + \mara_R \dfrac{\tilde{h}^2}{2} \tilde{\bnabla} \xi.
  \label{eq:lubrication_dimensionless_Q}
\end{equation}
\
\noindent Here, $\tilde{h} = 0.5 (1-\tilde{x}^2)$ is the dimensionless height profile, $\tilde{j} = 2/(\pi \sqrt{1-r^2})$ is the evaporation rate in the flat droplet limit and $\mara_R$ is the modified Marangoni number, defined as:
\
\begin{equation}
  \mara_R = \dfrac{R \partial_{\massfrac_A} \sigma}{D_0 \mu_0} \evap,
\end{equation}
\
We impose a homogeneous bulk correction for $\xi$, i.e., $\langle \xi \rangle = 0$, to allow for quasi-stationary solutions with a fixed compositional average. 
While the system of equations \eqref{eq:lubrication_dimensionless_div_Q}--\eqref{eq:lubrication_dimensionless_Q} can be solved analytically (see Appendix \ref{appA}), introducing a small amplitude $\epsilon$ of an azimuthal perturbation to the base solution makes the augmented system too complex for analytical solutions. 
Therefore, we employ numerical methods to determine the azimuthal stability of the quasi-stationary axisymmetric solutions and compare these results with those obtained using the full model described in \S\ref{sec:physical_model_stokes_droplet}.
As $\theta \rightarrow 0$, the results from the full model converge to the lubrication model, validating the latter (see figure \ref{fig:lubrication_validation}b).
Remarkably, the number of unstable modes $m$ is found to be consistent in both models. 
Interestingly, this number is larger for low $\theta$ values than for large $\theta$ values (see figure \ref{fig:lubrication_validation}a), i.e. a cascade of azimuthal instabilities sets in for decreasing $\theta$ at unexpectedly small negative Marangoni numbers.
By expanding the $\xi$ field into a sum of the base solution and a small amplitude perturbation in the direction of the corresponding eigenvector that triggers the instability of mode $m$, we can visualise the linear effects of the azimuthal instability for the lubrication model.

\subsection{Driving mechanism of the azimuthal instabilities}\label{sec:geometry_azimuthal_instability}

In the absence of bulk gradients, the driving mechanism of the Marangoni instabilities for low $\theta$ values is different from the one described in the two-dimensional box of \S\ref{sec:introduction}, yet it bears resemblance.

\begin{figure}
  \centering
  \includegraphics[width=1\textwidth]{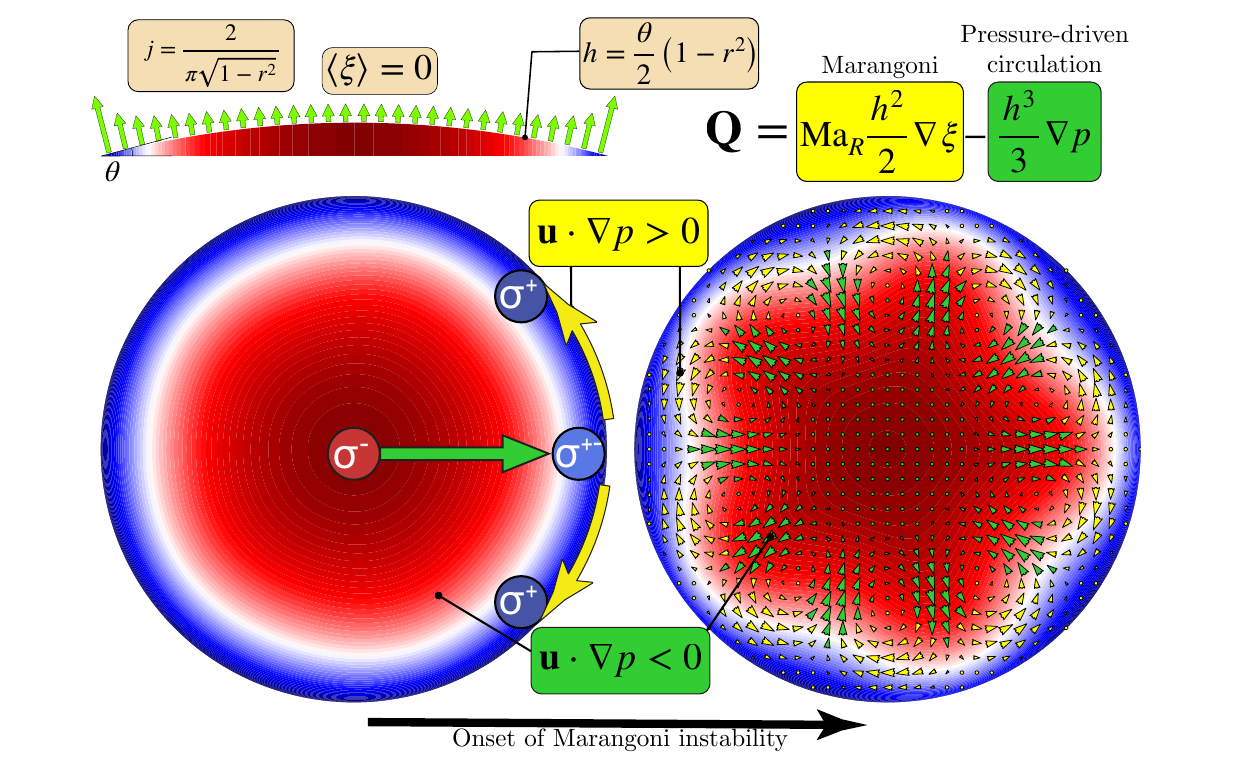}
  \caption{Side (left top) and top (left bottom) views of an evaporating droplet prior to the onset of the Marangoni instability.
  \reviewerC{The sign of the pressure gradients indicates whether the flow is dominated by Marangoni flow (positive pressure gradient) or whether it is pressure-driven flow (negative pressure gradient).}
  Should a disturbance locally reduce the surface tension at the contact line, it triggers an azimuthal Marangoni flow (yellow arrows) that transports fluid away from the perturbed region.
  This initiates pressure-driven return flow (green arrow) that attracts more fluid with lower surface tension from the centre of the droplet (greater-height region) towards the contact line (lower-height region).
  Ultimately, an azimuthal Marangoni instability is triggered (right).}
  \label{fig:flat_drop_ma_vs_p}
\end{figure}

For droplets with small $\theta$ and a negative Marangoni number ($\mara_R$), the larger evaporation rate at the contact line leads to an enhanced concentration of fluid with greater surface tension.
Should a disturbance locally reduce the surface tension at contact line, it triggers an azimuthal Marangoni flow that transports fluid away from the perturbed region.
\reviewerA{Continuity of the fluid} initiates a pressure-driven return flow, which attracts more fluid with lower surface tension towards the contact line (see figure \ref{fig:flat_drop_ma_vs_p}).
This creates a positive feedback loop, which is the driving mechanism of the azimuthal instabilities observed in droplets with low contact angle $\theta$.

Naturally, the same reasoning still holds for droplets with larger contact angle $\theta$.
However, in this latter case, the presence of bulk gradients becomes more pronounced, primarily driving the Marangoni instabilities.
A simple observation of the scaling of the different terms in equation \eqref{eq:lubrication_dimensionless_Q} reveals that the flow induced by $\mara_R$ scales with $h^2$, while the flow driven by pressure scales with $h^3$.
As a consequence, in areas where the height profile is minimal, the Marangoni flow prevails, but in areas with a significant height profile, pressure-driven flow takes precedence.
Thus, in droplets with a low $\theta$, the flow near the contact line is predominantly driven by Marangoni flow, whereas pressure-driven circulation dominates the centre (as illustrated by the arrows in figure \ref{fig:flat_drop_ma_vs_p}).

Therefore, the geometry plays a fundamental role in the driving mechanism of the onset of azimuthal instabilities.
In the next subsection, we propose a geometry where such instabilities can be obtained for positive-$\mara_R$ droplets (e.g.\ glycerol-water mixtures).

\subsection{Instabilities in positive Marangoni mixtures}\label{sec:instabilities_glycerol_water}

The insights gained from prior analyses elucidate that, in droplets with small $\theta$, if the fluid with larger surface tension has the lowest height profile, azimuthal instabilities can be triggered.
For positive-$\mara_R$ mixtures, the enhanced evaporation at the contact line leads to higher surface tension in the centre, which is the zone of larger height.
Thereby, positive-$\mara_R$ mixtures consistently exhibit stable axisymmetric solutions.
Any disturbance reducing the surface tension at the centre is promptly neutralised by a Marangoni flow in the opposite direction.

\begin{figure}
  \centering
  \includegraphics[width=1\textwidth]{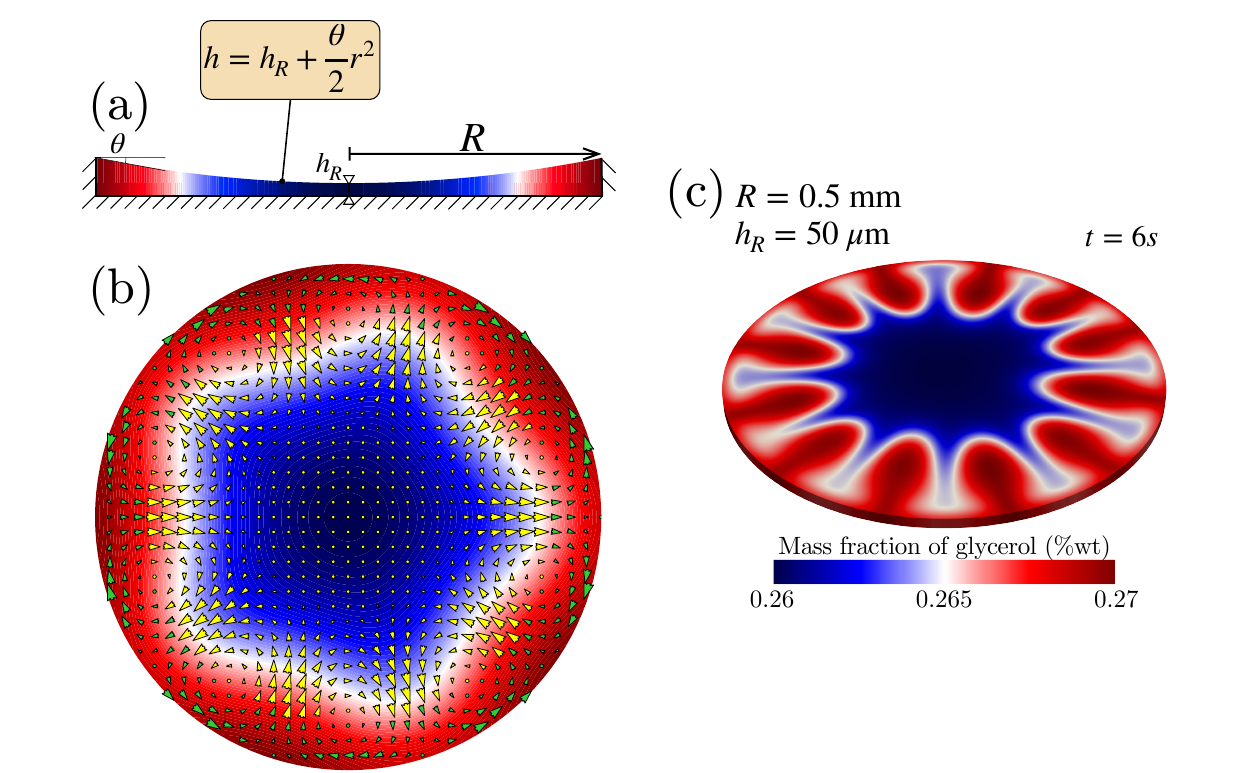}
  \caption{Side (a) and top (b) views of a droplet placed on a shallow pit. 
  Due to the inverted height profile, $\mara_R$ (yellow arrows) is predominant at the centre over the pressure-driven flow (green arrows), resulting in the onset of azimuthal Marangoni instabilities, depicted in (b).
  After enough evaporation time (\SI{6}{\second}), blobs emerge close to the contact line in realistic case of a $R=$\SI{0.5}{\milli \meter}, $\theta$=\SI{10}{\degree}, positive-$\mara_R$ droplet in a $h_R=$\SI{0.5}{\micro \meter} pit, with initial $\massfrac_\mathrm{glycerol}=$\SI{25}{\percent} (c).}
  \label{fig:glycerol_water_instability}
\end{figure}

Altering the scenario by placing the droplet within a shallow pit with a concave liquid-gas interface reverses the height profile, making the centre the lowest point (see figure \ref{fig:glycerol_water_instability}a).
This setup assumes that the depression is shallow enough for lubrication theory to hold and that the fluid remains pinned to the pit's sidewalls.
Under these conditions, a perturbation reducing surface tension at the centre instigates a Marangoni flow that shifts fluid azimuthally away from the disturbed area, thereby initiating an azimuthal Marangoni instability akin to the mechanism described in \S\ref{sec:geometry_azimuthal_instability}, albeit in a reversed direction (see figure \ref{fig:glycerol_water_instability}b).
This behaviour is unexpected when following the explanation provided in \S \ref{sec:introduction} for Marangoni instabilities in a two-dimensional box.
In that scenario, positive-$\mara$ mixtures cannot initiate Marangoni instabilities.

To simulate the evaporation dynamics of a realistic glycerol-water droplet with a radius of $R=$\SI{0.5}{\milli\meter}, $\theta$=\SI{10}{\degree}, in a pit of depth $h_R=$\SI{0.5}{\micro\meter}, starting with a $\massfrac_\mathrm{glycerol}=$\SI{25}{\percent}, we apply equations \eqref{eq:lubrication_h}--\eqref{eq:lubrication_Q} in a two-dimensional circular domain.
As evaporation progresses, blobs begin to form near the rim of the well due to the azimuthal Marangoni instability (see figure \ref{fig:glycerol_water_instability}c).
We emphasise that the pit must be shallow enough for the onset of instabilities, otherwise bulk gradients would stabilise the flow. While this finding is remarkable, a detailed investigation of the azimuthal Marangoni dynamics of droplets in such a well is beyond the scope of the work.

\color{orange}

  \subsection{Capillary effects in the Marangoni instability}\label{sec:capillary_effects}
  
\begin{figure}
    \centering
    \includegraphics[width=1\textwidth]{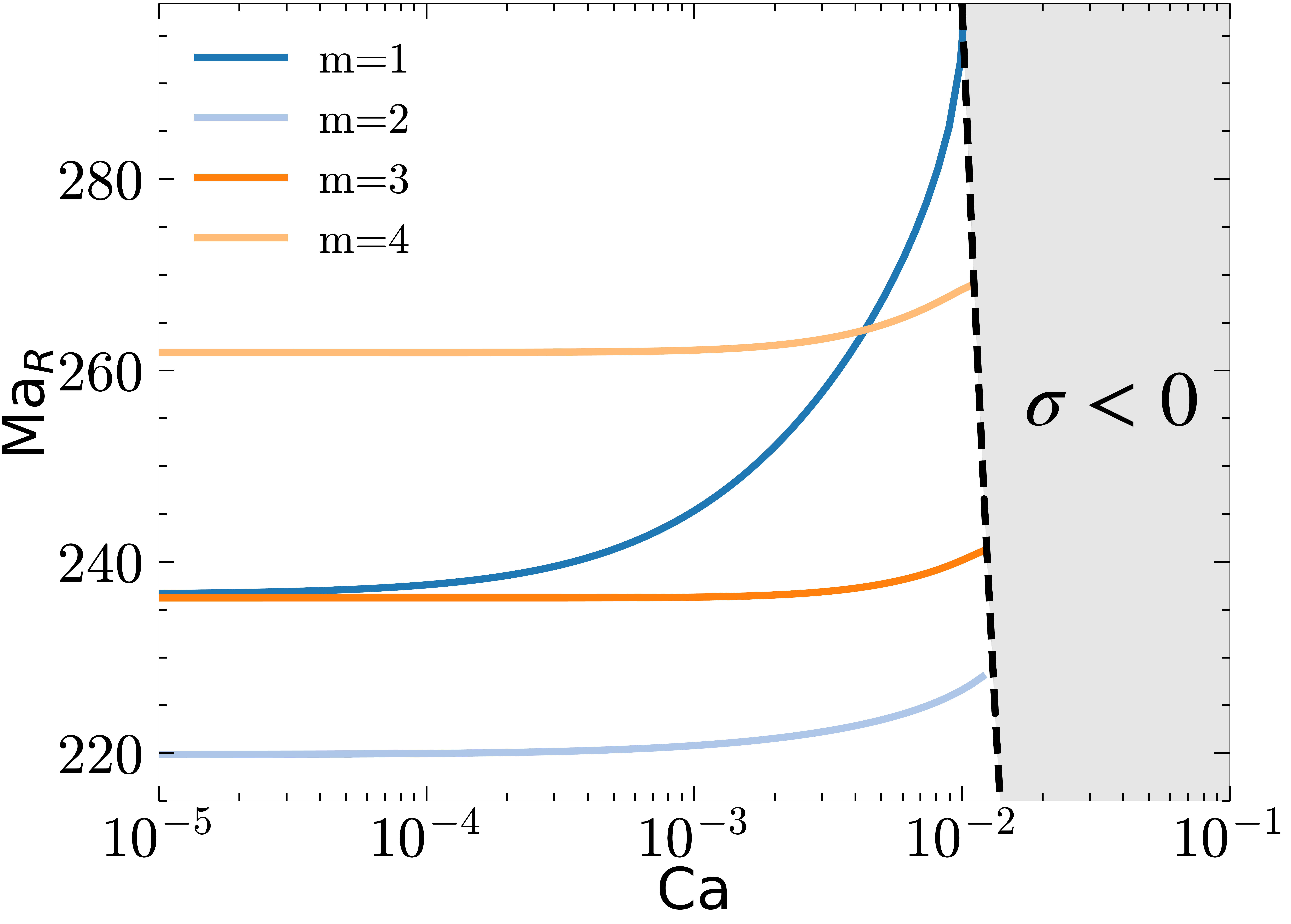}
    \caption{
      Critical Marangoni number ($\mara_R$) for the onset of azimuthal instabilities as a function of the capillary number $\mathrm{Ca}$ for wavenumbers $m=$ 1, 2, 3 and 4 and for contact angles of $10^\circ$ (a) and $20^\circ$ (b).
    The gray area, limited by the black dashed line, indicates where the surface tension is negative, rendering physically unrealistic solutions, therefore discarded from the analysis.
    The pink and yellow regions highlight the areas where the droplet maintains a nearly spherical-cap shape and where it takes a pancake shape, respectively.
    }
    \label{fig:non_zero_capillary}
  \end{figure}

  Until now, we assumed that the droplet retains a perfectly spherical-cap shape (parabolic shape in the lubrication limit) during evaporation, neglecting capillary effects on the onset of instabilities. 
  In this subsection, we explore non-zero capillary numbers to examine the influence of capillary effects, in particular shape deformations induced by surface tension gradients and Marangoni flow, on the onset of azimuthal instabilities. 
  Utilising the scaling introduced in \S\ref{sec:lubrication_numerical_method} and defining the capillary number as:
  \
  \begin{equation}
    \mathrm{Ca} = \dfrac{D_0 \mu_0}{\theta R \sigma_0},
  \end{equation}
  \
  \noindent we can obtain the nondimensionalised form of equations \eqref{eq:lubrication_h}--\eqref{eq:lubrication_Q}, namely
  \
  \begin{equation}
    \partial_{\tilde{t}} \tilde{h} + \tilde{\bnabla} \bcdot \tilde{\mathbf{Q}} = 0, 
    \label{eq:nonzeroCa_lubrication_dimensionless_h}
  \end{equation}
  \begin{equation}
    \tilde{p} + \bnabla \bcdot \left(\theta^2 (\mathrm{Ca}^{-1} + \mara_R \xi) \bnabla \tilde{h} \right) = 0,
    \label{eq:nonzeroCa_lubrication_dimensionless_p}
  \end{equation}
  \begin{equation}
    \partial_{\tilde{t}}(\xi \tilde{h}) + \tilde{\bnabla} \bcdot (\tilde{\mathbf{Q}} \xi) = \tilde{\bnabla} \bcdot (\tilde{h} \tilde{\bnabla} \xi) - \tilde{j},
    \label{eq:nonzeroCa_lubrication_dimensionless_y}
  \end{equation}
  \
  The definition of $\tilde{\mathbf{Q}}$ remains the same as in equation \eqref{eq:lubrication_dimensionless_Q}.
  Again, we neglect the ``coffee-stain effect'' in equation \eqref{eq:nonzeroCa_lubrication_dimensionless_h}, in the sense that we assume that the volume of the droplet does not change.
  By solving the system of equations \eqref{eq:nonzeroCa_lubrication_dimensionless_h}--\eqref{eq:nonzeroCa_lubrication_dimensionless_y} for various capillary numbers and contact angles of $10^\circ$ (a) and $20^\circ$ (b), we can identify the critical $\mara_R$ for the onset of azimuthal instabilities across different wavenumbers $m$ (see figure \ref{fig:non_zero_capillary}) and understand how the results of figure \ref{fig:lubrication_validation} are affected by shape deformations. 
  We ensure that the capillary number $\mathrm{Ca}$ is set such that the effective dimensionless surface tension, given in equation \eqref{eq:nonzeroCa_lubrication_dimensionless_p} by $\sigma = \theta^2 (\mathrm{Ca}^{-1} + \mara_R \xi$), remains positive.
  Beyond this limit, we discard the solutions as they are physically unrealistic, as depicted by the gray areas of subfigures \ref{fig:non_zero_capillary}(a) and (b), limited by the black dashed lines.

  The results indicate that, within the physical constraints of the capillary number (Ca), the critical Marangoni number ($\mara_R$) for the onset of azimuthal instabilities is significantly influenced by shape deformations. 
  For sufficiently large Marangoni numbers, the droplet adopts a pancake shape, as observed in both experimental and numerical studies \citep{diddens2017modeling, pahlavan2021evaporation, yang2023evaporation}.
  In each subfigure, we highlight the regions in the phase diagram where the droplet maintains a nearly spherical-cap shape (pink) and where it transitions to a pancake shape (yellow).
  The critical transition between these shapes was determined by identifying the set of parameters ($\mara_R$, $\mathrm{Ca}$) at which the vertically-averaged pressure at the highest point of the droplet along the axis of symmetry becomes zero.
  
  An increase in the modulus of the critical Marangoni number is observed across all azimuthal wavenumbers $m$, with a more pronounced effect for $m=1$ and for $\theta=10^\circ$, where $|\mara_R|$ increases from approximately $200$ to $10^5$ as $\mathrm{Ca}$ rises from 0 to around $10^{-3}$. 
  Interestingly, the effects of shape deformations are less pronounced for droplets with larger contact angles, see subfigure \ref{fig:non_zero_capillary}(b).
  In evaporating droplets with positive Marangoni number and with unpinned contact line, the Marangoni flow can so strongly drive fluid towards the apex that the droplet is contracted \citep[``Marangoni contraction'', ][]{karpitschka2017marangoni}.
  Similarly, a pinned droplet with negative Marangoni number can form a pancake-shape \citep{pahlavan2021evaporation, yang2023evaporation}, where strong Marangoni flow drives fluid from the apex of the droplet towards the contact line, causing deformation and an increase in the height profile at the contact line. 
  The microscopic contact angle also becomes higher for pancake droplets.
  This agrees with the observations of our model: instabilities for flat droplets start at the rim and the higher the contact angle, the larger $|\mara_R|$ is required for instabilities.
  
\color{black}

  \section{Chaotic behaviour of negative Marangoni evaporating droplet}\label{sec:chaotic_behaviour}
  
Until this section, we carefully characterised the observed flow as ``seemingly'' chaotic. Here, we utilise the zero-$\mathrm{Ca}$ lubrication model from \S\ref{sec:lubrication_numerical_method} within a circular domain to accurately capture the radial and azimuthal flow features of flat droplets, i.e.\ the full nonlinear long-term behaviour beyond the onset of instability. 
Our objective is to characterise the complex spatio-temporal dynamics of flat evaporating droplets with negative $\mara_R$, specifically to determine whether the system exhibits chaotic behaviour.

In a chaotic system, small changes in the initial conditions exponentially grow in time. 
In dissipative systems, the distance between two phase space trajectories with slightly different initial conditions remains bounded within a strange attractor \citep{strogatz2018nonlinear}. 
To investigate this dynamical system, we monitor the evolution of a small perturbation $\mathbf{U}(t) + \delta \mathbf{U}(t)$ relative to the unperturbed solution $\mathbf{U}(t)$, under the assumption that both will eventually converge to the same strange attractor. 
We analyse the linear growth of $\delta \mathbf{U}(t)$ around $\mathbf{U}(t)$.
Despite the system's nonlinearities, the long-term dynamics are expected to exhibit exponential growth or decay, represented as $\delta \mathbf{U}(t) \sim e^{\lambda t}$, provided  that $\delta U$ stays tiny, i.e. nonlinear terms do not take action yet.
In an $N$-dimensional dynamical system, there are typically $N$ Lyapunov exponents $\lambda_i$, with $i = 1, \ldots, N$ \citep{strogatz2018nonlinear}. 
If at least one of these exponents is positive, it indicates that a random perturbation will generally grow over time, signifying chaotic behaviour.

To calculate the Lyapunov exponents, we start with one or more initial random perturbations $\delta \mathbf{U}_i(0)$, for $i \le N$.  
Here, $N$ is the number of degrees of freedom of the discretised composition $\xi$, which contains the only time derivative. 
If $i > 1$, we perform Gram-Schmidt orthogonalization \citep{wiesel1993continuous, christiansen1997computing} at each step to ensure that slower-growing perturbations remain orthogonal to the fastest-growing ones. 
The Lyapunov exponents are then determined from their definition
\
\begin{equation}
  \lambda_i = \lim_{\reviewerC{t}\rightarrow\infty} \lim_{|\delta \mathbf{U}_i^0|\rightarrow 0} \dfrac{1}{t} \log{\dfrac{|\delta \mathbf{U}_i (t)|}{|\delta \mathbf{U}_i^0|}}\,.
\end{equation}
\
\noindent In practice, however, we calculate the exponential growth over a finite dimensionless averaging time $\bar{t}$, since the infinite time limit cannot be realised in numerical calculations. 
We use $\bar{t} = \mathrm{min}(t,20)$ to bypass the initial transient dynamics of the evaporative process, which would otherwise skew the Lyapunov exponent calculations and necessitate computationally more expensive simulations.

To avoid a completely transient system, we consider a droplet with a fixed volume $V$ and a constant average concentration field $\langle \xi \rangle$. 
Although this assumption may not be entirely realistic during the temporal evolution of an evaporating flat droplet, our goal is to provide insights into the deterministic nature of the flow field within the droplet at a specific time when the droplet has a defined volume and average concentration.

\begin{figure}
  \centering
  \includegraphics[width=1\textwidth]{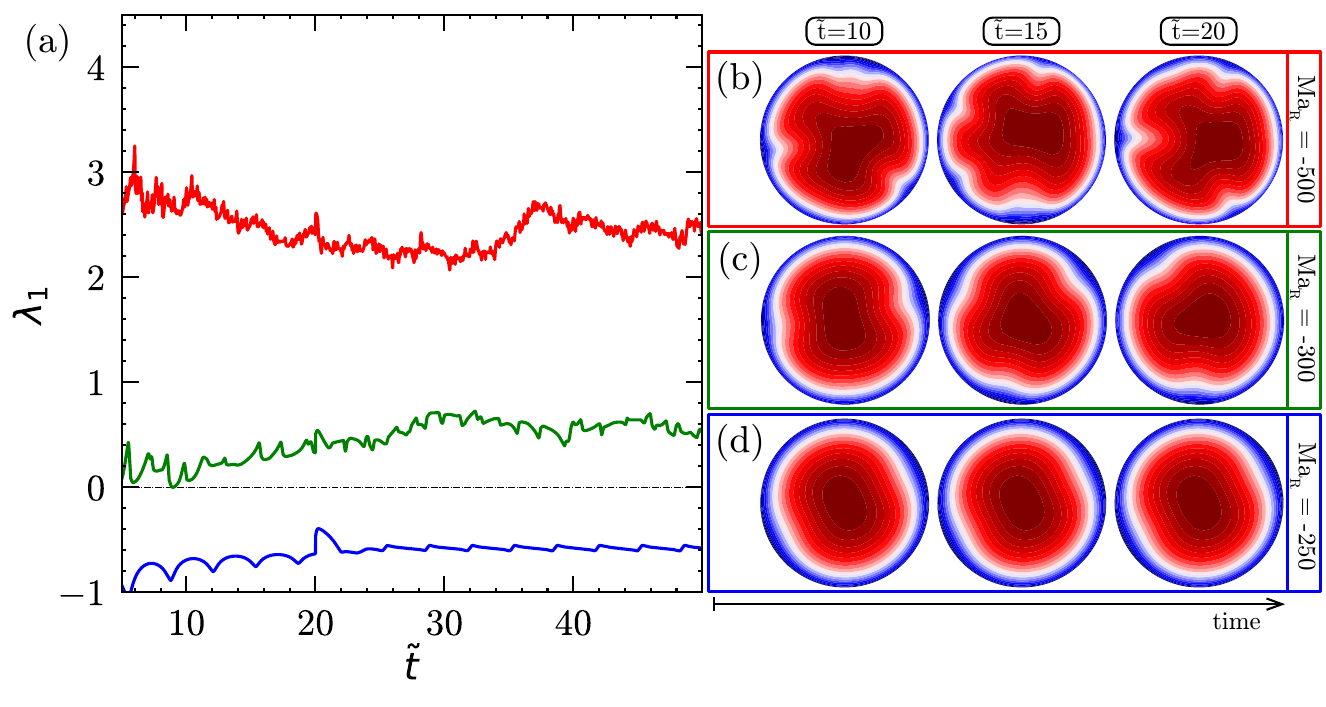}
  \caption{Largest Lyaponov exponent for $\mara_R = -500$ (red), $\mara_R = -300$ (green), and $\mara_R = -250$ (blue) (a), plotted over dimensionless time $\tilde{t}$.
  The black dotted line indicates the transition to chaotic behaviour, i.e.\ when the largest Lyapunov exponent becomes positive.
  The $\xi$ profile at $\tilde{t} =$ 10, 15, and 20 for $\mara_R = -500$ (b), $\mara_R = -300$ (c), and $\mara_R = -250$ (d).
  }
  \label{fig:chaotic_behaviour}
\end{figure}

For both $\mara_R = -500$ and $\mara_R = -300$, the largest Lyapunov exponent is positive, as shown by the red and green curves in figure \ref{fig:chaotic_behaviour}(a). 
The $\xi$ profile changes significantly over time, displaying irregular and aperiodic patterns, as seen in figures \ref{fig:chaotic_behaviour}(b) and (c). 
\textcolor{orange}{In supplementary movie 4, we show the evolution of the composition field over time and the calculation of the Lyaponov exponents at each instance, for $|\mara_R|=10000$.}
Our calculations indicate that higher $|\mara_R|$ values result in more pronounced chaotic behaviour, evidenced by larger Lyapunov exponents.
When $|\mara_R|$ is reduced to 250, the largest Lyapunov exponent becomes negative, as shown by the blue curve in figure \ref{fig:chaotic_behaviour}(a). 
The solutions appear to stabilise over time, as illustrated in figure \ref{fig:chaotic_behaviour}(d), and a dominant $m=2$-unstable mode is observed in the resulting quasi-stationary $\xi$ profile.

For a critical $|\mara_R|$ between 250 and 300, the largest Lyapunov exponent approaches zero, indicating a transition to chaotic behaviour. 
By applying the linear stability analysis method outlined in \S\ref{sec:physical_model_stokes_droplet} to the quasi-stationary solutions of the two-dimensional system in the lubrication limit, we can pinpoint the critical $\mara_R$ at which this transition occurs. 
We identify a subcritical Hopf bifurcation at $\mara_R = -268.69$, signalling the onset of chaotic dynamics and limiting the system's deterministic behaviour. 
This relatively low critical $\mara_R$ value aligns with experimental observations of violent and random flow fields in evaporating droplets with negative $\mara_R$ \citep{machrafi2010benard,diddens2017evaporating}. 
For instance, a water-ethanol droplet with a \SI{1}{\milli\meter} radius, a \SI{10}{\degree} contact angle, and \SI{10}{\wtpercent} ethanol, evaporating at ambient conditions and relative humidity of \SI{50}{\percent}, can easily reach an order of magnitude $\mara_R \sim -10^7$. 
Our findings confirm that these observations are indicative of chaotic nonlinear behaviour rather than merely the superposition of linearly unstable modes.

\color{orange}
\section{Limitations and applicability of results}\label{sec:limitations}

While rich in physical insights, the results presented in this work are based on minimal models that simplify the intricate dynamics of evaporating binary droplets. 
In reality, droplets containing a highly volatile component, such as water--ethanol mixtures, evaporate rapidly, leading to significant transient effects. 
The evolving composition alters the physical properties of the liquid phase, while evaporative cooling induces temperature gradients that drive additional Marangoni flows. 
The impact of the latter depends on factors such as the droplet's contact angle, the thermal conductivity of the substrate, and the latent heat of evaporation of the components. 
Further complexities can arise due to Raoult's law, Stefan's flow and natural convection in the gas phase, both of which can substantially influence real systems.  

Incorporating all these effects into a mathematical model is challenging due to the numerous parameters involved. 
Our approach is to begin with a simplified system, isolating the onset of instabilities driven by solutal Marangoni effects, with the aim of extrapolating some conclusions to more complex cases. 
To achieve this, we focus on regimes with a low evaporation number, where many of these effects can be neglected. 
This can be realised experimentally by placing the droplet in a controlled environment where the far-field vapour concentrations of both components remain close to equilibrium, minimising transient effects and ensuring a nearly constant average concentration field, $\langle \xi \rangle$.  

Even in this simplified system, thermal effects may still be relevant \citep{karapetsas2012convective}. 
The thermal field typically evolves more rapidly than the composition field, as indicated by the small Lewis number, $\mathrm{Le} = D/\alpha$, where $\alpha$ is the thermal diffusivity. 
Consequently, temperature gradients can develop swiftly, inducing thermal Marangoni flows that may temporarily enslave solutal Marangoni flows. 
However, solutal gradients are usually much stronger than thermal ones. 
This suggests that over time, solutal Marangoni effects can become dominant.  
In other words, one can state that the solutal Marangoni number $\mara$ significantly exceeds the thermal Marangoni number, which we define as 
\begin{equation}
  \mara_T = \frac{V^{1/3} \partial_{T}\sigma \rho_0 c_{p,0}}{k_0 \mu_0} \mathrm{Le}^{-1} \mathrm{Ev}_\text{tot},
\end{equation}
\begin{gather}
  \text{with } \mathrm{Ev}_\text{tot} = \frac{D_A^\text{vap} (c_A^\text{eq} - c_A^\infty) - D_B^\text{vap} (c_B^\text{eq} - c_B^\infty)}{\rho_0 D_0},
\end{gather}
\noindent where $k_0$ is the thermal conductivity, $c_{p,0}$ is the specific heat capacity, and $\partial_{T}\sigma$ is the temperature derivative of the surface tension evaluated at the average temperature of the droplet. The total evaporation number, $\mathrm{Ev}_\text{tot}$, is consistent with the definition of \citet{diddens2021competing}.

Thermal effects could be minimised by selecting an appropriate substrate and controlling ambient conditions. 
In principle, these could be completely eliminated by adjusting the far-field vapour concentration such that one component evaporates at the same rate as the other condenses, i.e.\ $j_A = - j_B$. 
In other words, the total evaporation number $\mathrm{Ev}_\text{tot}$ must vanish. 
In this scenario, temperature gradients disappear, the droplet volume remains constant, and solutal gradients persist—making our model directly applicable. While such cases are rare in practical applications, we explore a slight relaxation of this condition in \S\ref{sec:example_real_system}.  

The minimal models presented here constitute a first step towards understanding the onset of Marangoni instabilities in evaporating droplets. 
The results should be interpreted as a qualitative guide to the factors influencing instability development in systems with negative Marangoni numbers.  

\subsection{Slowly evaporating axisymmetric water-ethanol droplet}\label{sec:example_real_system}

We consider a slowly evaporating axisymmetric water--ethanol droplet. 
The mass fractions of water, ethanol, and air in the far-field of the gas phase are adjusted at each step to maintain constant $\mathrm{Ev}=0.001$ and $\mathrm{Ev}_\text{tot}=0.005$. 
These values are chosen to ensure that the $\mara$ number remains within the limit in which stable axisymmetric quasi-stationary solutions exist and that the thermal Marangoni number $\mara_T$ remains at least one order of magnitude lower than the solutal Marangoni number $\mara$, while still allowing for the droplet to evaporate.

\begin{figure}
  \includegraphics[width=1\textwidth]{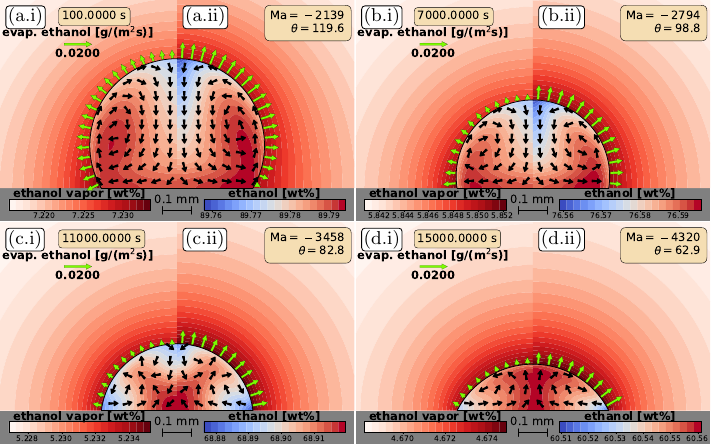}
  \caption{Slowly evaporating water--ethanol droplet with an initial contact angle of $\theta = 120^\circ$ and an initial volume of $V_0 = \SI{0.1}{\micro \liter}$. 
  The contact line is pinned to the substrate. 
  The far-field mass fractions of ethanol and water vapours are adjusted such that the evaporation number $\mathrm{Ev} = 0.001$ and the total evaporation number $\mathrm{Ev}_\text{tot}=0.005$. 
  Different stages of the droplet's evolution are shown at $t = \SI{100}{\second}$ (a), $t = \SI{7000}{\second}$ (b), $t = \SI{11000}{\second}$ (c), and $t = \SI{15000}{\second}$ (d). 
  Each panel shows the mass fraction of ethanol within the droplet and the mass fraction of ethanol in the vapour phase. 
  Results are presented for both the full transient simulations on the left of each panel (i) and the corresponding quasi-stationary solutions on the right (ii).}
  \label{fig:example_real_system}
\end{figure}

The droplet has an initial contact angle of $\theta = 120^\circ$ and an initial volume of $V_0 = \SI{0.1}{\micro \liter}$. 
The contact line is pinned. 
The substrate has a thickness of \SI{0.1}{\milli\meter}, it has the material properties of borosilicate, as in many experiments \citep[e.g.][]{deegan1997capillary, deegan2000contact}, and it has air beneath it. 
Thermal effects are incorporated into the transient model, as well as the effects of Raoult's law, Stefan's flow and natural convection in the gas phase. 
At each time step, the solutal Marangoni number $\mara$ and the contact angle are calculated. 
These are then used as input parameters for the quasi-stationary model of \S\ref{sec:physical_model_stokes_droplet}. 
As Figure \ref{fig:ma_vs_theta} shows, the quasi-stationary solutions exhibit hysteresis. 
To directly compare the results of the full transient simulations with the quasi-stationary solutions, we use the results of the transient model as an initial condition for the quasi-stationary calculations. 
The results are shown in Figure \ref{fig:example_real_system}.
Supplementary movies 1 and 2 shows the full transient evolution of the droplet and its comparison to the quasi-stationary model (1 for composition, 2 for velocity magnitude). Supplementary movie 3 shows the water vapour concentration in the gas phase and temperature field in the droplet and gas phases in the transient simulations.

Despite the absence of transient effects, thermal effects, and the effects of Raoult's law, Stefan's flow and natural convection in the gas phase, the quasi-stationary solutions are in good agreement with the full transient simulations. 
In the transient model, the lower temperature at the apex of the droplet induces a thermal Marangoni flow that drives fluid from the apex to the rim. 
It also decreases the vapour pressure of ethanol at the apex, leading to lower evaporation at the apex -- opposite to what is seen in the quasi-stationary model. 
Nevertheless, the flow is predominantly driven by solutal Marangoni effects, as demonstrated by the strong agreement of flow direction and compositional profile between the two models. 
In (a) and (b), the flow is in the $\rimtoapex$ direction, in (c), it is in the $2V$ direction, and in (d), it is in the $\apextorim$ direction, showing the presence of all three quasi-stationary solution regimes.

\color{black}

\section{Conclusions}\label{sec:outlook}

Evaporating multicomponent droplets with a negative Marangoni number (e.g.\ water-ethanol droplets) can exhibit interfacial instabilities \citep{pearson1958convection,sternling1959interfacial}.
These instabilities arise due to the sign of the solutal Marangoni number, i.e.\ whether surface tension increases or decreases with the mass fraction of the most volatile component \citep{gelderblom2022evaporation}.
In the latter case, as the most volatile component evaporates, surface tension becomes greater at the liquid-gas interface compared to the surface tension corresponding to the liquid composition in the bulk.
Any perturbation that locally reduces the surface tension at the interface, will trigger a Marangoni flow that transports fluid away from the disturbed region.
In a sufficiently thick bulk domain, continuity forms a vertical return flow, enhancing the perturbation by drawing more fluid with low surface tension towards the perturbed region.
Remarkably, our calculations reveal that the onset of instabilities occurs at a much lower critical $|\mara_\mathrm{cr}|$ for droplets with low $\theta$ than for droplets with a more significant bulk domain.

To investigate this phenomenon, we first identified the quasi-stationary axisymmetric solution regimes — vortex from rim to apex ($\rimtoapex$), vortex from apex to rim ($\apextorim$), and two vortices (2V) — near $\theta = 90^\circ$ as a function of $\mara$, and subsequently explored the full $\mara$-$\theta$ phase space using a minimal model.
We then assessed the azimuthal stability of each regime individually.
In the $\rimtoapex$ regime, the axisymmetric solution remains stable for $|\mara|$ values below a critical $m=1$ bifurcation but becomes unstable for $m=1$ above this threshold.
By further increasing $|\mara|$, more bifurcation curves for different $m$ values are crossed -- up to $m=8$.
All quasi-stationary solutions in the 2V regime were $m=1$-unstable.
The $\apextorim$ regime revealed a multitude of bifurcations for droplets with too low contact angle $\theta$, with several unstable modes $m$, contradicting the intuitive expectation that Marangoni instabilities would be suppressed -- or at least reduced -- in the absence of a thick bulk domain.

We then applied a simplified lubrication model to explore the onset of azimuthal instabilities in the small $\theta$ limit.
Although we derived an analytical axisymmetric solution, the perturbed fields were too complex to be solved analytically, leading us to employ numerical methods to uncover the mechanism driving these instabilities.
We found that the higher evaporation rate at the contact line enhances the concentration of fluid with higher surface tension.
A disturbance locally reducing surface tension at the contact line initiates an azimuthal Marangoni flow, which removes fluid from the perturbed region and triggers a pressure-driven azimuthal return flow.
This flow attracts more low-surface-tension fluid towards the contact line, creating a positive feedback loop that drives the instabilities.
By analysing the scales of the Marangoni ($h^2$) and pressure-driven ($h^3$) flows in equation \eqref{eq:lubrication_dimensionless_Q}, we demonstrated that the droplet's height profile plays a crucial role in the onset of these instabilities.
If the region of lowest height coincides with the highest surface tension, azimuthal instabilities can occur.
Thereby, placing a glycerol-water droplet in a shallow pit, reversing the height profile, can induce such instabilities, opposed to a glycerol-water droplet on a flat substrate, which does show azimuthal instabilities. 
This finding stresses the importance of the geometry on the interfacial instabilities.

It is important to note that these results were obtained using a minimal model in which the droplet was assumed to retain a spherical-cap shape during evaporation -- an assumption that may not hold in real systems \citep{diddens2017modeling, pahlavan2021evaporation, yang2023evaporation}.
We also investigated the influence of non-zero capillary numbers on the onset of azimuthal instabilities.
\reviewerD{The results suggest that, depending on the contact angle, the critical Marangoni number $\mara_R$ for azimuthal instabilities can be significantly affected by shape deformations.
Depending on the control parameters, the droplet either takes a nearly spherical-cap shape or some pancake-like shape, which can delay the onset of instabilities.}
We explored the chaotic behaviour of evaporating droplets with negative $\mara_R$ using a two-dimensional lubrication model.
We determined that the critical $\mara_R = -268.69$ marks the onset of chaotic dynamics, which aligns with experimental observations of violent flow fields in such droplets.

\reviewerB{Lastly, we discuss the limitations and applicability of our results.
We acknowledge that the models presented here are minimal and do not capture the full complexity of evaporating binary droplets.
In real systems, transient effects, thermal effects, and the effects of Raoult's law and Stefan's flow in the gas phase can significantly influence the dynamics.
However, by focusing on regimes with low evaporation numbers, we can neglect many of these effects and provide a qualitative guide to the onset of instabilities driven by solutal Marangoni effects.
We also present an example of a slowly evaporating water-ethanol droplet, where we incorporate thermal effects and the effects of Raoult's law and Stefan's flow in the gas phase.
The quasi-stationary solutions were in agreement with the full transient simulations.}

In conclusion, we have presented a detailed analysis of the onset of azimuthal instabilities in evaporating droplets with negative $\mara_R$ number.
By isolating the solutal Marangoni effects, we demonstrated that these instabilities can occur even when thermal effects are neglected.
Our results highlight the fundamental role of the height profile in driving instabilities for flat droplets.
Additionally, we showed that the chaotic flows observed in evaporating droplets with negative $\mara_R$ are due to inherent chaotic dynamics rather than the superposition of linearly unstable modes.

This work opens new avenues for further research, such as investigating the evaporation of droplets with positive Marangoni number in shallow pits, which could extend the work of \citet{d2023effect}.
Incorporating more complex effects, such as a comprehensive evaporation model or ``coffee-stain flow'', could provide a more realistic representation of the system and facilitate direct comparisons with experimental results. Also, secondary bifurcations could be identified and Lyapunov exponents for high contact angle droplets could be extracted from full three-dimensional simulations. This, however, constitutes a numerically expensive and demanding task.

From a broader perspective, our analysis demonstrates (i) how incredibly rich the flow regimes of evaporating binary droplets is and (ii) that nevertheless, it is, at least in part and with justified approximations, accessible to mathematical analysis.

\section*{Declaration of Interests}
The authors report no conflict of interest.

\section*{Acknowledgements}
This work was supported by an Industrial Partnership Programme, High Tech Systems and Materials (HTSM), of the Netherlands Organisation for Scientific Research (NWO); a funding for public-private partnerships (PPS) of the Netherlands Enterprise Agency (RVO) and the Ministry of Economic Affairs (EZ); Canon Production Printing Netherlands B.V.; and the University of Twente.

The authors would like to thank Alvaro Marin for valuable discussions and feedback on the manuscript.

\appendix
\section{Analytical axisymmetric solution in lubrication limit and zero-$\mathrm{Ca}$}\label{appA}

The system of equations \eqref{eq:lubrication_dimensionless_div_Q} to \eqref{eq:lubrication_dimensionless_Q} can be solved analytically for the quasi-stationary axisymmetric base solution in the limit of small $\theta$ and zero capillary number.
The existence of an axisymmetric flow requires the absence of z-averaged velocity, or equivalently, $\mathbf{Q^0} = 0$.
Here, the tilde was omitted for simplicity and the superscript $0$ denotes the axisymmetric base solution.
In consequence, there will be no net advective transport of the composition, which simplifies equation \eqref{eq:lubrication_dimensionless_y} to a Poisson equation with a source term ($j$).
The analytical solution can thus be expressed as:
\
\begin{equation}
  \mathbf{Q}^0 = 0
  \label{eq:axisymmetric_solution_Q},
\end{equation}
\begin{equation}
  \xi^0 = \dfrac{-6r^2 + 12 \artanh(\sqrt{1-r^2}) + 12 \log(r) - 5}{3 \pi \theta},
  \label{eq:axisymmetric_solution_xi}
\end{equation}
\begin{equation}
  p^0 = \dfrac{12 \mara_R}{\pi \theta^2} \left( \dfrac{1}{\sqrt{1-r^2}} - \log (1 + \sqrt{1-r^2}) \right),
  \label{eq:axisymmetric_solution_p}
\end{equation}
\
\noindent where $r$ is the radial coordinate in the axisymmetric coordinate system. 
Despite the presence of a logarithm in the $\xi^0$ solution, it remains well-defined at $r = 0$. 
This is due to the $\log$-singularity being counterbalanced by the equivalent divergence of $\artanh$.
However, as anticipated, the pressure exhibits a divergence at the contact line when the Marangoni number is non-zero.

When we introduce a small amplitude $\epsilon$ perturbation to expand each field $f$ into $f = f^0 + \epsilon f^m e^{im\phi + \lambda t}$, the simplification of $\mathbf{Q} = 0$ no longer applies. 
This is because the divergence of the expanded $\mathbf{Q}$ field is no longer zero.
As a result, the advective transport of the composition is non-zero, yielding complex analytical expressions for the perturbed fields.
Given these complexities, it was not possible to derive an analytical solution for the perturbed fields in the lubrication limit.

\bibliographystyle{myjfm}
\bibliography{bibliography}

\end{document}